\documentclass[prl,twocolumn,aps,longbibliography,superscriptaddress,notoc,groupedaddress]{revtex4-1}
\usepackage{bm}
\usepackage{graphicx}
\usepackage{amssymb}
\usepackage{amsmath}
\usepackage{eufrak}
\usepackage{color}
\usepackage[utf8]{inputenc}
\usepackage[english]{babel}
\usepackage{ulem}
\usepackage[unicode=true,colorlinks=true,citecolor=blue,urlcolor=blue]{hyperref}

\renewcommand{\d}{\mathrm d}
\renewcommand{\emph}{\textit}
\newcommand{\braket}[1]{\left\langle #1 \right\rangle}

\usepackage{chngcntr}
\newcommand{\enquote}{}

\newcommand{\nix}[1]{}

\renewcommand{\section}[1]{\textit{#1}---}

\begin{document}

\title{Dynamic Valley Polarization in Moir\'e Quantum Dots}

\author{D.~S.~Smirnov}
\email[Electronic address: ]{smirnov@mail.ioffe.ru}
\affiliation{Ioffe Institute, 194021 St. Petersburg, Russia}

\begin{abstract}
  We describe the effects of the crossing of the exciton levels in moir\'e quantum dots in external magnetic field.
  We demonstrate that the unpolarized light can create significant dynamic valley polarization of interlayer excitons in the conditions where the Zeeman energy is much smaller than the thermal energy.
  The angular momentum is gained by the excitons from the nuclear spin fluctuations.
  In the $n$-type moir\'e quantum dots for the typical experimental parameters, the unpolarized light can lead to the complete valley polarization of the resident electrons.
\end{abstract}

\maketitle{}

\section{Introduction}The recent discovery of the twisted two-dimensional heterostructures opened a new era in semiconductors physics~\cite{Tran_2020}. For transition metal dichalcogenides (TMDCs) bilayers, a small twist angle leads to the long range moir\'e potential, which significantly changes the excitonic properties~\cite{seyler2019signatures,tran2019evidence,jin2019observation,alexeev2019resonantly} and can lead to many interesting strongly correlated many body states~\cite{tang2020simulation,regan2020mott,shimazaki2020strongly,wang2020correlated}.

Despite the weak van-der-Waals interaction between the monolayers, the lattice reconstruction is significant at small twist angles and creates enlarged areas with the energetically favorable stacking~\cite{weston2020atomic,doi:10.1021/acsnano.0c04832}. This results in the strong localizing potential for the charge carriers, so for the low temperatures and weak excitation powers the moir\'e quantum dots (MQDs) emerge~\cite{Yue1701696,brotons2020spin,shabani2021deep}.


The strong spin-orbit interaction lifts the spin degeneracy at the edges of the band gap in TMDCs monolayers, so the lowest Kramers degenerate states can be labeled by the valley pseudospin. The valley manipulation is the main goal of the valleytronics of MQDs. In this field, the single photon emission, optical orientation, fine structure splitting of the trions, nearest neighbour interactions, and single charge carrier thermal spin polarization were already demonstrated experimentally~\cite{jin2019identification,Baekeaba8526,liu2021signatures,brotonsgisbert2021moiretrapped,baek2021optical}.
The further progress is anticipated in the direction of optical valley initialization, manipulation and readout in MQDs. In this Letter we put forward a novel valley polarization mechanism for MQDs: dynamic valley polarization. It is based on the weak electron-hole exchange interaction in the interlayer excitons and the large exciton oscillator strength. It takes place under unpolarized optical excitation at small magnetic fields and at the high temperatures as compared with the fine structure splitting. It is robust against inhomogeneous broadening of the optical resonances.

The dynamic valley polarization appears resonantly in the external magnetic at the crossing of the intravalley and intervalley excitonic states. The degree of dynamic valley polarization of both excitons and resident charge carriers can reach 100\% for the realistic MQDs parameters.

\begin{figure}
  \centering
  \includegraphics[width=\linewidth]{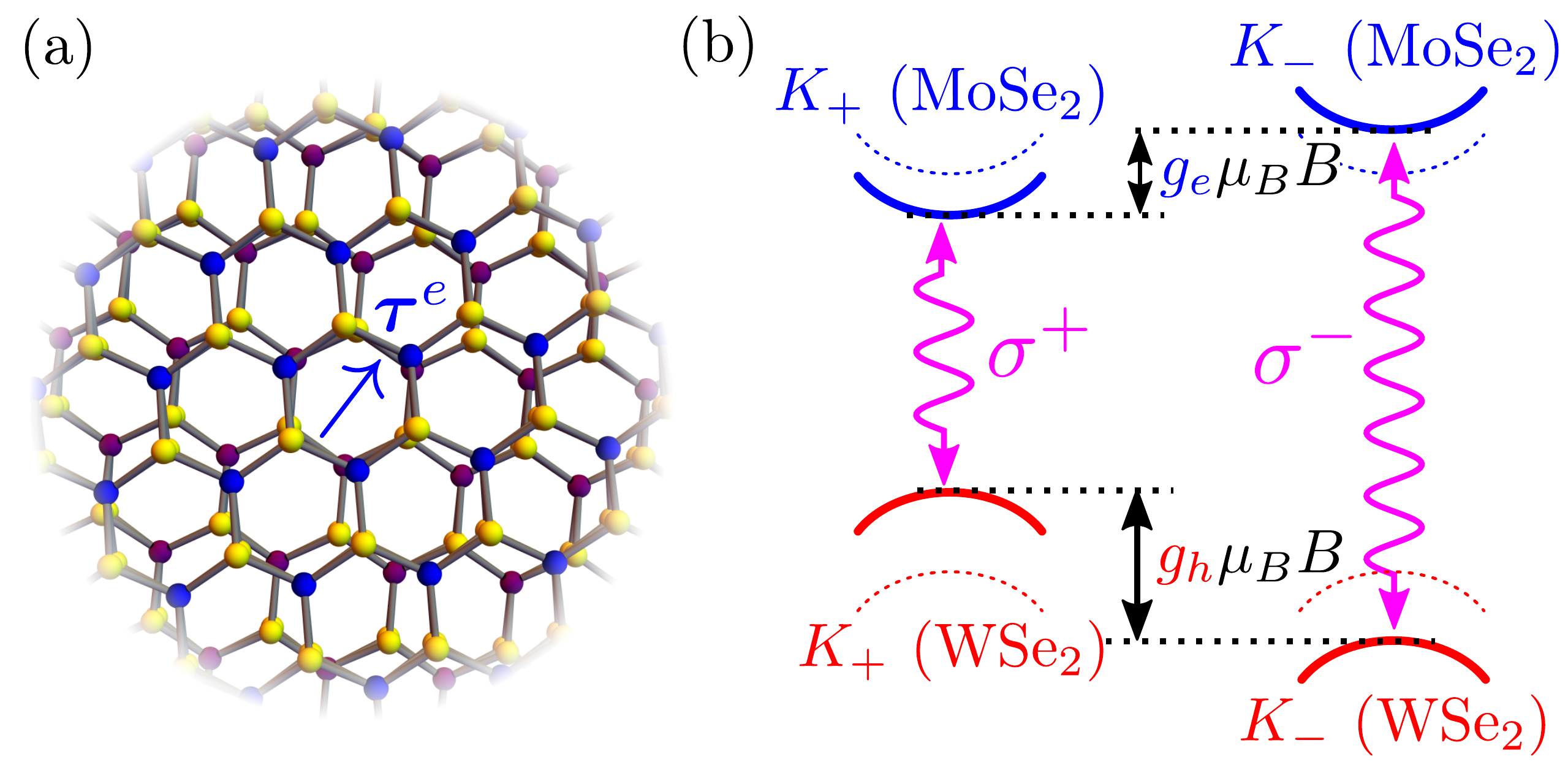}
  \caption{(a) Sketch of a MoSe$_2$/WSe$_2$ MQD with 2H stacking. W, Mo, and Se atoms are shown by purple, blue and yellow balls, respectively. (b) Electron and hole valley states under consideration in a MQD. The thin dotted and thick solid acrs correspond to zero and finite external magnetic field.}
  \label{fig:intro}
\end{figure}

To be specific we focus on 2H MoSe$_2$/WSe$_2$ MQDs, as shown in Fig.~\ref{fig:intro}(a), however the presented concept is equally valid for all types of MQDs. In our case, the ground states of electrons and holes are in MoSe$_2$ and WSe$_2$ monolayers, respectively. We assume the fast electron and hole energy relaxation as compared with the exciton recombination rate, so that we can consider only the lowest conduction subband and the uppermost valence subband of the corresponding layers, as shown in Fig.~\ref{fig:intro}(b). In these subbands the spins $\bm s^{e,h}$ and valley pseudospins $\bm\tau^{e,h}$ of electron and hole, respectively, are locked as $\tau_z^{e,h}=-s_z^{e,h}$~\cite{forg2019cavity}, where $z$ is the axis normal to the MQD. So the exciton dynamics can be described using only the valley pseudospins $\bm\tau^{e}$ and $\bm\tau^h$.

The spin and valley relaxation of excitons in TMDC monolayers can be related with the interplay between scattering of excitons and electron-hole exchange interaction or with the electron-phonon interaction~\cite{Kioseoglou2012,PhysRevB.89.201302,PhysRevB.89.205303,MX2Review}. However, for the individual non-interacting electrons and holes localized in MQDs in weak magnetic fields, these mechanisms are inefficient due to the absence of orbital motion and vanishing of electron-phonon interaction at zero frequency~\cite{MX2_Avdeev}. This is typical for the QDs made of the conventional semiconductors, where the spin relaxation in weak magnetic fields is driven by the hyperfine interaction~\cite{book_Glazov}. Thus we assume that the valley relaxation in MQDs is also determined by the interaction with the host lattice nuclei.


\section{Neutral MQD}
\begin{figure}
  \centering
  \includegraphics[width=0.9\linewidth]{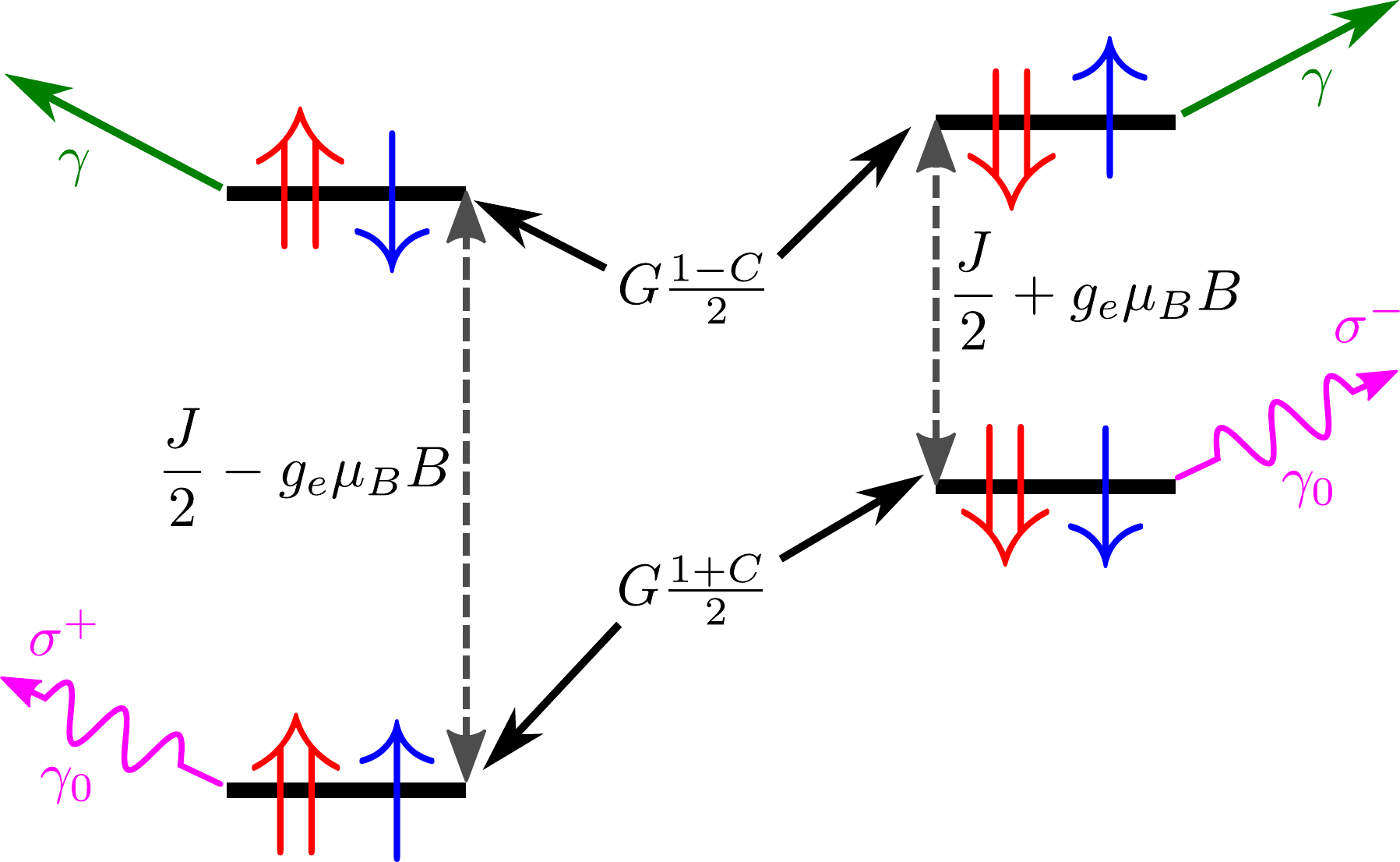}
  \caption{Exciton levels, splittings between them (gray dashed arrows), generation (black solid arrows), radiative (magenta wavy arrows) and nonradiative (green solid arrows) recombination channels. Red double and blue single arrows denote the hole and electron valley pseudospins $\tau_z^h=\pm1/2$ and $\tau_z^e=\pm1/2$, respectively; $g_e<0$ is assumed.}
  \label{fig:levels}
\end{figure}

Firstly, let us consider the exciton generation in a MQD by unpolarized light. The four exciton states can be labeled by $\tau_z^e=\pm1/2$ and $\tau_z^h=\pm1/2$ corresponding to $K_+$ and $K_-$ valleys, respectively. The exciton Hamiltonian accounting for the exchange interaction, Zeeman effect and the hyperfine interaction has the form
\begin{equation}
  \mathcal H=-J\tau_z^e\tau_z^h+g_e\mu_B(B\tau_z^e+\bm B_N^e\bm\tau^e)+g_h\mu_B(B+B_N^h)\tau_z^h,
\end{equation}
where $J$ is the exchange interaction constant related to the electron-hole short-range exchange interaction~\cite{PhysRevLett.115.176801,nl_exch_1,Deilmann_2019,doi:10.1021/acs.nanolett.0c00633,Yu_exchange}, $\mu_B$ is the Bohr magneton, $B$ is the external magnetic field directed along the $z$ axis, $g_{e,h}=g_v^{e,h}-g_s^{e,h}$ are the electron and hole $g$-factors in the subbands under consideration, respectively, with $g_{v,s}^{e,h}$ being the corresponding valley and spin $g$-factors~\cite{Wang_2015,DurnevUFN}, $\bm B_N^{e,h}$ are the random nuclear fields acting on electron and hole valley pseudospins in the MQD, respectively. The exciton fine structure at $\bm B_N^{e,h}=0$ is shown in Fig.~\ref{fig:levels}.

For holes, $\bm B_N^h$ is directed along the $z$ axis~\cite{MX2_Avdeev}. So it can not lead to the hole valley flips and does not play a role in exciton dynamics. For electrons, by contrast, $\bm B_B^e$ can have all three components. This nuclear field is described by the Gaussian distribution function~\cite{MX2_Avdeev}
\begin{multline}
  \label{eq:F}
  \mathcal F(\bm B_N^e)=\frac{1}{2(\sqrt{\pi}\Delta_B)^3}\\\times\exp\left[-\frac{(B_{N,x}^e)^2+(\Omega^e_{N,y})^2}{\Delta_B^2}-\frac{(B^e_{N,z})^2}{(2\Delta_B)^2} \right],
\end{multline}
where $\Delta_B$ is the typical fluctuation of the nuclear field. Assuming the number of nuclear spins to be large, $\bm B_N^{e,h}$ can be consider as frozen, i.e. static~\cite{merkulov02}. Below we consider the limit of weak hyperfine interaction as compared with the exchange interaction, $\Delta_B\ll J/(|g_e|\mu_B)$. In the same time, the Zeeman splitting can be of the order of $J$ as shown in Fig.~\ref{fig:levels}, so it can even change the order of the exciton levels.

We consider unpolarized optical exciton generation. Typically electron-hole pairs are photogenerated through the direct optical transitions of individual monolayers because of the higher oscillator strength of the intralayer excitons. Then electron or hole tunnels to another layer with lower energy. This process can preserve some correlation degree between valleys of photogenerated electron and hole. Thus the generation rates of intra- and intervalley excitons are $G(1\pm C)/2$, respectively, where $G$ is an average generation rate and $C$ is a valley correlation degree. The exciton generation is shown in Fig.~\ref{fig:levels} by the black solid arrows.

The intravalley excitons can recombine radiatively with the rate $\gamma_0$ emitting $\sigma^+$ or $\sigma^-$ light depending on the valley~\cite{Yu_2018,forg2019cavity}, as shown in Fig.~\ref{fig:levels}. In the same time, they are in the spin triplet state, so they have lower energy then the intervalley excitons, which corresponds to $J>0$. For the intervalley excitons we consider the possibility of the nonradiative recombination with the small rate $\gamma\ll\gamma_0$.

In external magnetic field the exciton levels shift and at $B\approx\pm B_{\rm ex}$, where $B_{\rm ex}=J/(2|g_e|\mu_B)$, the intravalley and intervalley excitons can be efficiently mixed by the interaction of electron and nuclei. In this case, in the second order in the hyperfine interaction strength we find the recombination rates of the intervalley excitons due to the mixing with the intravalley excitons with $\tau_z^h=\pm1/2$:
\begin{equation}
  \gamma_\pm=\frac{B_{N,x}^2+B_{N,y}^2}{4(B_{\rm ex}\pm B)^2+B_0^2}\gamma_0,
\end{equation}
where
\begin{equation}
  \label{eq:B0}
  B_0=\hbar\gamma_0/(|g_e|\mu_B)
\end{equation}
describes the broadening of the intravalley exciton levels due to the radiative recombination.

In the steady state, the numbers of excitons are given by the ratio of the corresponding generation and recombination rates. Thus we obtain
\begin{subequations}
  \label{eqs:exc}
  \begin{equation}
    n_{\Downarrow\downarrow}=n_{\Uparrow\uparrow}=\frac{G(1+C)}{2\gamma_0},
  \end{equation}
  \begin{equation}
    n_{\Downarrow\uparrow}=\frac{G(1-C)}{2(\gamma+\gamma_-)},
    \qquad
    n_{\Uparrow\downarrow}=\frac{G(1-C)}{2(\gamma+\gamma_+)},
  \end{equation}
\end{subequations}
where $\Uparrow(\Downarrow)$ and $\uparrow(\downarrow)$ denote the hole and electron valley pseudospin, respectively, see Fig.~\ref{fig:levels}.

\begin{figure}
  \centering
  \includegraphics[width=0.9\linewidth]{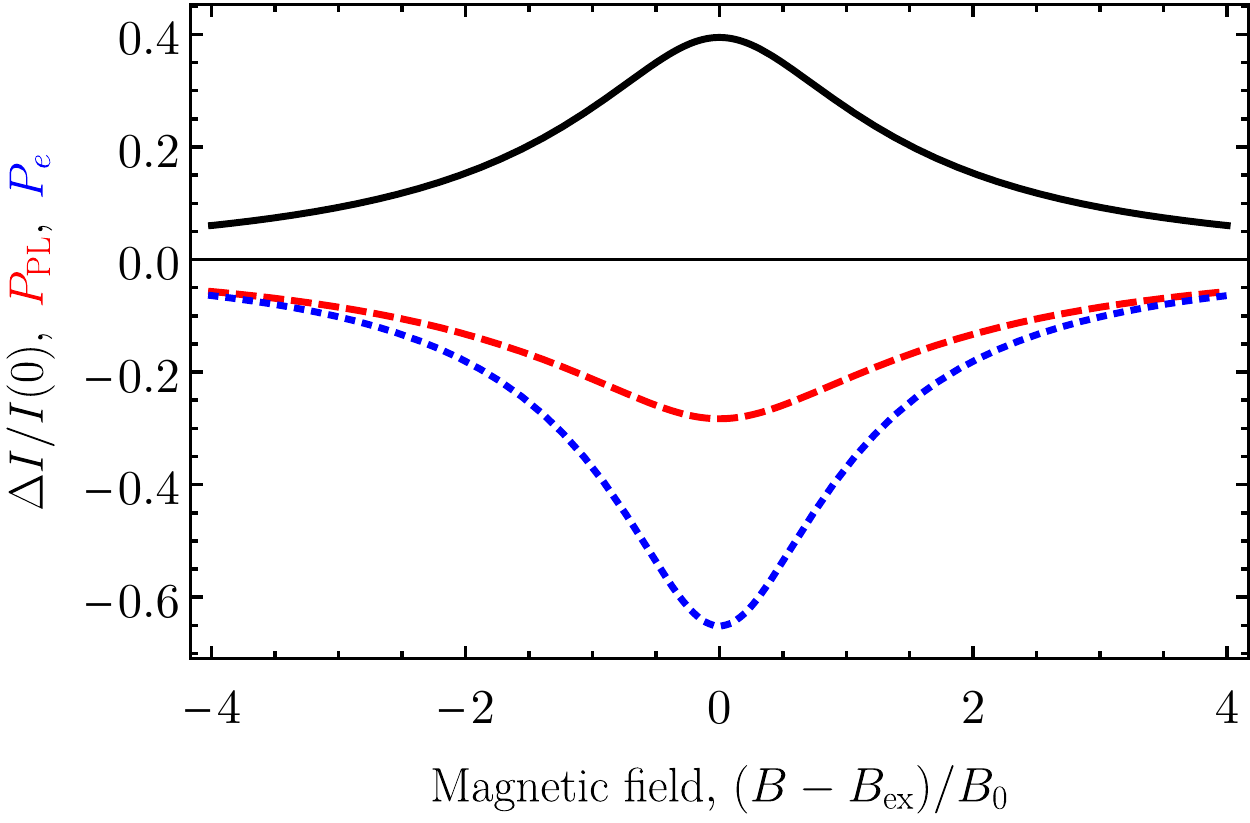}
  \caption{Relative change of the intensity, $\Delta I/I(0)$, (black solid curve) and polarizations of PL, $P_{\rm PL}$, (red dashed curve) and electron valley, $P_e$, (blue dotted curve) for neutral MQD as functions of the magnetic field in the vicinity of $B_{\rm ex}$ for $C=0$, $\gamma/\gamma_0=10^{-3}$, and $\Delta_B/B_0=0.1$.
  }
  \label{fig:exc_pol}
\end{figure}

The total intensity of the photoluminescence (PL) of excitons in MQD is given by
\begin{equation}
  I=\gamma_0(n_{\Uparrow\uparrow}+n_{\Downarrow\downarrow})+\braket{\gamma_+n_{\Uparrow\downarrow}+\gamma_-n_{\Downarrow\uparrow}},
\end{equation}
where the angular brackets denote averaging over the distribution function~\eqref{eq:F}. Here the first two terms describe the recombination of the intravalley excitons. The two latter terms describe the recombination of intervalley excitons assisted by the electron-nuclear spin flips. It takes place at the crossings of exciton levels at $|B|\approx B_{\rm ex}$. As a result, the intensity of the PL resonantly increases by $\Delta I=I(B)-I(0)$. This is shown in Fig.~\ref{fig:exc_pol} by the black solid curve. Here the valley correlations are neglected, $C=0$, so the intravalley and intervalley excitons are generated with the same rates. At the crossing of the levels the intervalley excitons with the corresponding spin can recombine radiatively, which increases the PL intensity by up to $50$\%.

When the PL intensity increases, it gets also circularly polarized, because only one type of intervalley excitons mix with the intravalley excitons and emit circularly polarized light. The polarization degree is given by
\begin{equation}
  P_{\rm PL}=\braket{\gamma_+n_{\Uparrow\downarrow}-\gamma_-n_{\Downarrow\uparrow}}/I;
\end{equation}
it is shown in Fig.~\ref{fig:exc_pol} by the red dashed curve. One can see that it is negative and in the optimal conditions it can reach $1/3$.

In Fig.~\ref{fig:exc_pol} we use the dimensionless parameters, which can be associated with the following realistic values: $1/\gamma_0=10$~ns~\cite{PhysRevLett.126.047401,Jiang2018}, $1/\gamma=10~\mu$s~\cite{Jiang2018}, and $\hbar/(|g_e|\mu_B\Delta_B)=100$~ns~\cite{MX2_Avdeev}. The spatial separation of electron and hole suppresses the radiative recombination by $\sim10^3$ times, and we assume the same suppression for the exchange interaction. As a result we can estimate $B_{\rm ex}$ as $20$~mT for $J=6~\mu$eV~\cite{Yu_exchange} with $g_e=-5.6$~\cite{PhysRevB.101.235408,PhysRevResearch.2.033256,PhysRevLett.124.226402}. We note that at the level crossing the thermal polarization of the PL is positive and is only a few percent at $2$~K. Thus it can be easily separated in the experiment from the dynamic polarization, which is negative and takes place resonantly at $B\approx B_{\rm ex}$. 

\begin{figure}[t]
  \centering
  \includegraphics[width=\linewidth]{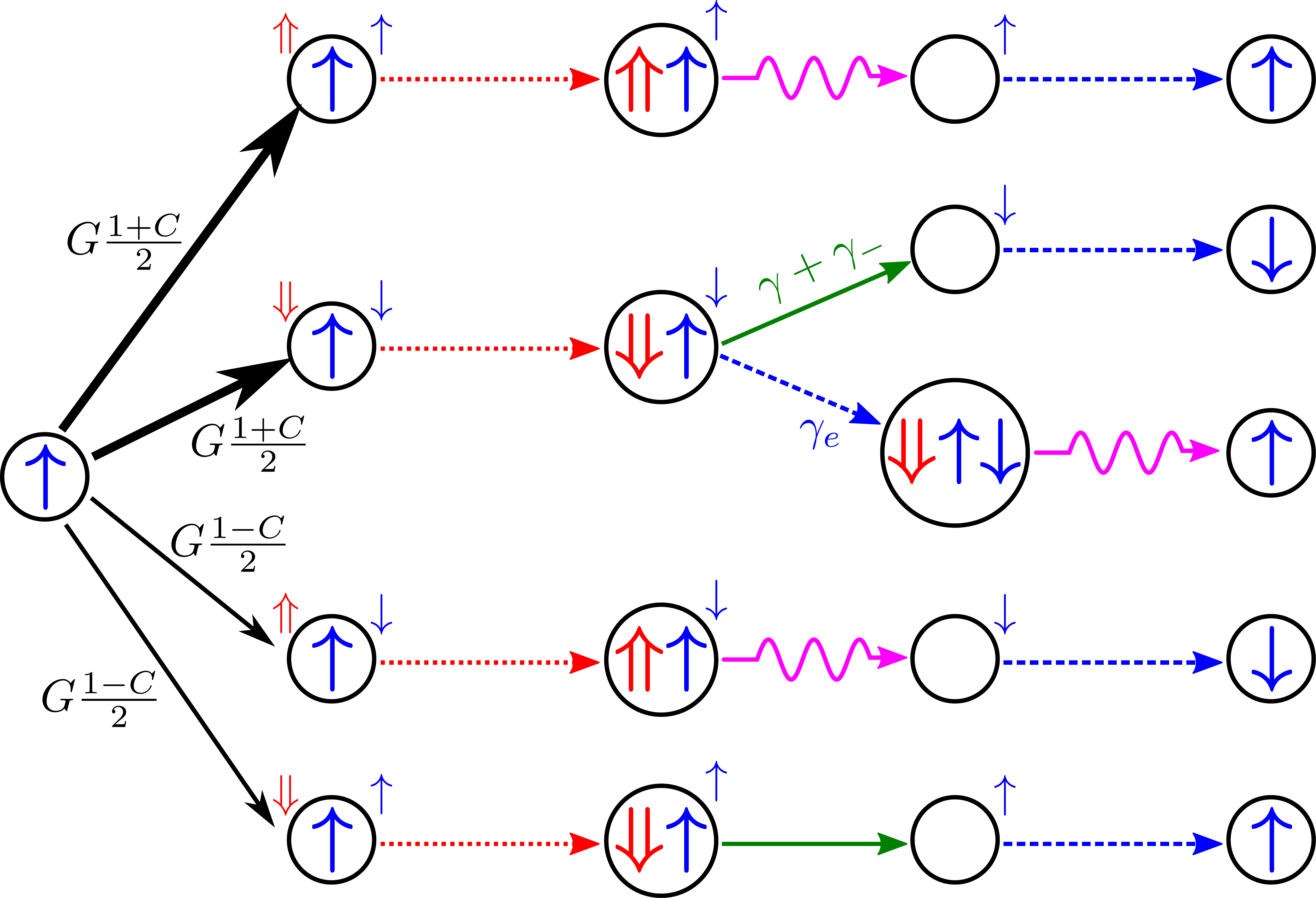}
  \caption{Principle of the electron dynamic valley polarization. Arrows in circles and close to them show electron and hole spins in the MQD and in its vicinity, respectively. Thick upper and thin lower black arrows show the generation of intra- and intervalley electron-hole pairs, respectively. Red dotted and blue dashed arrows show hole and electron tunneling in the MQD, respectively. Green and wavy magenta arrows show nonradiative and radiative exciton recombinations, respectively.}
  \label{fig:electron_polarization}
\end{figure}

In the same conditions, the electron in exciton also gets valley polarized. Its polarization degree is
\begin{equation}
  P_e=\frac{\braket{n_{\Uparrow\uparrow}+n_{\Downarrow\uparrow}-n_{\Downarrow\downarrow}-n_{\Uparrow\downarrow}}}{\braket{n_{\Uparrow\uparrow}+n_{\Downarrow\uparrow}+n_{\Downarrow\downarrow}+n_{\Uparrow\downarrow}}},
\end{equation}
which is also negative, as shown in Fig.~\ref{fig:exc_pol} by the blue dotted curve. Interestingly, it can reach 100\% when the lifetime of one type of the intervalley excitons is much longer than that of another.

We note that because of the time reversal symmetry, the change of the intensity $\Delta I$ is an even function of the magnetic field, while the PL and electron valley polarizations are odd. The valley correlation $C$ suppresses the effects of the levels crossing, because for $C=1$ only the intravalley excitons are generated, which produce unpolarized PL with the constant intensity.

\section{Charged MQDs}Now let us consider in a similar way $n$-type MQDs charged with single resident electrons. We assume that the additional electron-hole pairs are generated optically somewhere close to the given MQD, but not directly inside it. We again allow for the valley correlation degree $C$, as shown in Fig.~\ref{fig:electron_polarization}. The Coulomb repulsion prevents photoelectron from tunneling in the MQD, so the photogenerated hole tunnels first, and an exciton forms in the MQD. This exciton can recombine exactly as we described in the previous section. After recombination the photogenerated electron tunnels to the MQD conserving its valley. Additionally, we assume that the photogenerated electron can tunnel to the MQD if it has an opposite spin to the electron in exciton. In this case, the trion forms, and then recombines radiatively leaving behind a single electron in the MQD. We assume the electron tunneling rate $\gamma_e$ to be much smaller than the radiative recombination rate $\gamma_0$, but can be comparable with $\gamma$, see Fig.~\ref{fig:electron_polarization}. 

The resident electron valley dynamics described above can be summarized by the following kinetic equations:
\begin{multline}
  \frac{\d n_{\uparrow/\downarrow}}{\d t}=\pm G\frac{1-C}{2}(n_\downarrow-n_\uparrow)\\
  \pm G\frac{1+C}{2}\left(\frac{\gamma+\gamma_+}{\gamma_e+\gamma+\gamma_+}n_\downarrow-\frac{\gamma+\gamma_-}{\gamma_e+\gamma+\gamma_-}n_\uparrow\right),
\end{multline}
where $n_{\uparrow/\downarrow}$ are the numbers of electrons with $\tau_z^e=\pm1/2$, respectively, and we neglect the slow electron valley relaxation~\footnote{This assumption is valid for $\gamma_e\gg G\gg(|g_e|\mu_B\Delta_B)/\hbar$.}.

\begin{figure}[t]
  \centering
  \includegraphics[width=0.9\linewidth]{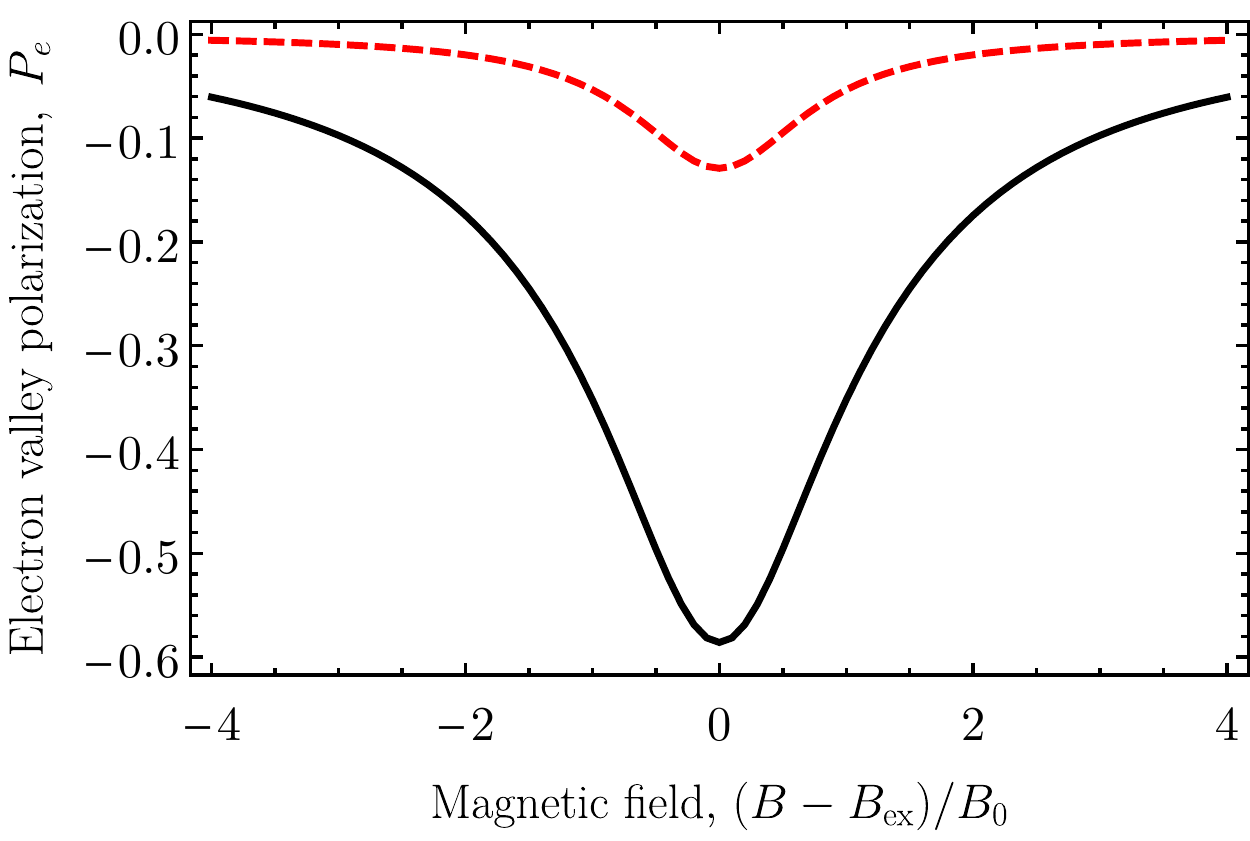}
  \caption{Resident electron dynamic valley polarization as a function of the magnetic field for $C=0$ (red dashed curve) and $C=1$ (black solid curve) with $\gamma_e/\gamma_0=10^{-2}$ and the other parameters the same as in Fig.~\ref{fig:exc_pol}.
  }
  \label{fig:resident}
\end{figure}

These equations in analogy with Eqs.~\eqref{eqs:exc} yield the steady state occupancies $n_{\uparrow/\downarrow}$. From them we obtain the electron valley polarization degree $P_e=\braket{n_\uparrow-n_\downarrow}/\braket{n_\uparrow+n_\downarrow}$, which is shown in Fig.~\ref{fig:resident}. It is negative (note the assumption of $g_e<0$) and appears resonantly at the crossing of the exciton levels.

Qualitatively this polarization takes place due to the electron valley flip channel shown by the upper green arrow in Fig.~\ref{fig:electron_polarization}. At resonance, it allows electron to flip its valley from $\tau_z^e=+1/2$ to $-1/2$, but not in the opposite direction due to the absence of the levels crossing for the exciton in the opposite valley.

Neglecting the valley correlations $C$ and nonradiative recombination $\gamma$, the polarization degree is $P_e=\braket{(\gamma_+-\gamma_-)/(2\gamma_e+3\gamma_++3\gamma_-)}$. Its absolute value does not exceed $1/3$, see the red dashed curve in Fig.~\ref{fig:resident}.

The valley correlations increase the dynamic electron valley polarization degree, as shown by the black curve in Fig.~\ref{fig:resident}. 
For $C=1$ and $\gamma\ll\gamma_e$ at $B\approx B_{\rm ex}$ the polarization degree is
\begin{equation}
  \label{eq:P_e}
  P_e=-\braket{\frac{\gamma_-}{2\gamma+\gamma_-}}.
\end{equation}
One can see, that if the electron spin flips dominate over the nonradiative recombination, $\gamma_-\gg\gamma$, the electron valley polarization reaches $100$\%.


\section{Discussion}
The most important result of this work is the dynamic valley polarization of the resident electrons in the charged MQDs, which can be easily realized experimentally in the state of the art MQDs. This mechanism is very efficient and has no analogues in the previous works. We stress that the dynamic valley polarization is different from the optical orientation and thermal polarization, because it does not require polarized or resonant excitation and takes place at the large temperatures as compared with the fine structure splittings.

The valley polarization of the resident electrons can be probed either directly through the PL circular polarization or independently using the valley-induced Kerr rotation of a linearly polarized probe light~\cite{Hsu2015,Yang2015nat}. The latter can be strongly enhanced at the frequencies of direct optical transitions of the individual monolayers. We note that the valley polarization of the resident charge carriers can be preserved for a few microseconds~\cite{Dey2017,Jin893}.

For transient excitons the dynamic valley polarization mechanism has some similarities with the magnetic circular polarization effect~\cite{Ivchenko2018}. However it is unrelated with the unequal thermal population of the excitonic levels and involves the electron-nuclear spin flips~\cite{PhysRevLett.125.156801,shamirzaev2021dynamic}. 

Along with the electron valley polarization, the nuclear spins can also get polarized in the same conditions, if their relaxation time is long enough. In MoSe$_2$ the hyperfine interaction is the strongest with Se atoms~\cite{MX2_Avdeev} and if their spins get polarized, they effectively enhance the external magnetic field. Thus the shift of the maximum of the dynamic valley polarization to smaller fields with increase of the excitation power is possible.

The hyperfine interaction in TMDC monolayers was not directly evidenced experimentally yet. However, its fingerprint is the non monoexponential spin relaxation, which can be suppressed by a weak magnetic field~\cite{PRC}. Such features were already observed~\cite{PhysRevX.6.021024,PhysRevMaterials.5.044001} indicating the presence of static random effective magnetic fields, which may be related to the host lattice nuclear spins.



As a possible alternative to the hyperfine interaction, one can use a tilted external magnetic field. In an isolated TMDC monolayer the translational invariance forbids valley mixing by the homogeneous magnetic field. But in moir\'e structures the translational invariance is broken, so one can expect a small transverse $g$-factors of the order of the ratio of the lattice constant and the moir\'e period. Thus the transverse (in-plane) components of the magnetic field lead to the mixing of intra- and intervalley excitons similarly to the hyperfine interaction and also produce the dynamic valley polarization.


If the energy relaxation is not fast enough, the upper subband of the conduction band can be significantly populated in addition to the lowest one. In this case our considerations remain qualitatively valid, and we expect the dynamic valley polarization of the order of unity. However, since the $g$-factors in the two conduction subbands $g_{c.b.}$ and $g_{c.b.+1}$ are different, the dynamic polarization can take place at the two different magnetic fields, corresponding to the two crossings of the exciton levels. Assuming the exchange interaction constant to be the same for the both bands, the ratio of the two resonant magnetic fields is equal to $g_{c.b.}/g_{c.b.+1}\approx5.6$ for MoSe$_2$~\cite{PhysRevB.101.235408,PhysRevResearch.2.033256,PhysRevLett.124.226402}.

\textit{In conclusion} we have studied the effects of the crossing of the exciton levels in MQDs in external magnetic field. 
It leads to the increase of the PL intensity and to its circular polarization.
We predict the dynamic valley polarization of excitons and of resident charge carriers in MQDs. This is a novel valley polarization mechanism, which takes place under unpolarized optical excitation and is robust against inhomogeneous broadening of the optical resonance and increase of the temperature. 
The degree of the dynamic valley polarization can reach 100\%.

\begin{acknowledgments}
  I thank \href{https://orcid.org/0000-0003-4462-0749}{M. M. Glazov} for fruitful and stimulating discussions. This work was partially supported by the Russian Science Foundation Grant No. 21-72-10035.
\end{acknowledgments}


\begin{thebibliography}{51}%
\makeatletter
\providecommand \@ifxundefined [1]{%
 \@ifx{#1\undefined}
}%
\providecommand \@ifnum [1]{%
 \ifnum #1\expandafter \@firstoftwo
 \else \expandafter \@secondoftwo
 \fi
}%
\providecommand \@ifx [1]{%
 \ifx #1\expandafter \@firstoftwo
 \else \expandafter \@secondoftwo
 \fi
}%
\providecommand \natexlab [1]{#1}%
\providecommand \enquote  [1]{``#1''}%
\providecommand \bibnamefont  [1]{#1}%
\providecommand \bibfnamefont [1]{#1}%
\providecommand \citenamefont [1]{#1}%
\providecommand \href@noop [0]{\@secondoftwo}%
\providecommand \href [0]{\begingroup \@sanitize@url \@href}%
\providecommand \@href[1]{\@@startlink{#1}\@@href}%
\providecommand \@@href[1]{\endgroup#1\@@endlink}%
\providecommand \@sanitize@url [0]{\catcode `\\12\catcode `\$12\catcode
  `\&12\catcode `\#12\catcode `\^12\catcode `\_12\catcode `\%12\relax}%
\providecommand \@@startlink[1]{}%
\providecommand \@@endlink[0]{}%
\providecommand \url  [0]{\begingroup\@sanitize@url \@url }%
\providecommand \@url [1]{\endgroup\@href {#1}{\urlprefix }}%
\providecommand \urlprefix  [0]{URL }%
\providecommand \Eprint [0]{\href }%
\providecommand \doibase [0]{http://dx.doi.org/}%
\providecommand \selectlanguage [0]{\@gobble}%
\providecommand \bibinfo  [0]{\@secondoftwo}%
\providecommand \bibfield  [0]{\@secondoftwo}%
\providecommand \translation [1]{[#1]}%
\providecommand \BibitemOpen [0]{}%
\providecommand \bibitemStop [0]{}%
\providecommand \bibitemNoStop [0]{.\EOS\space}%
\providecommand \EOS [0]{\spacefactor3000\relax}%
\providecommand \BibitemShut  [1]{\csname bibitem#1\endcsname}%
\let\auto@bib@innerbib\@empty
\bibitem [{\citenamefont {Tran}\ \emph {et~al.}(2020)\citenamefont {Tran},
  \citenamefont {Choi},\ and\ \citenamefont {Singh}}]{Tran_2020}%
  \BibitemOpen
  \bibfield  {author} {\bibinfo {author} {\bibfnamefont {K.}~\bibnamefont
  {Tran}}, \bibinfo {author} {\bibfnamefont {J.}~\bibnamefont {Choi}}, \ and\
  \bibinfo {author} {\bibfnamefont {A.}~\bibnamefont {Singh}},\ }\bibfield
  {title} {\enquote {\bibinfo {title} {Moir{\'{e}} and beyond in transition
  metal dichalcogenide twisted bilayers},}\ }\href {\doibase
  10.1088/2053-1583/abd3e7} {\bibfield  {journal} {\bibinfo  {journal} {2D
  Materials}\ }\textbf {\bibinfo {volume} {8}},\ \bibinfo {pages} {022002}
  (\bibinfo {year} {2020})}\BibitemShut {NoStop}%
\bibitem [{\citenamefont {Seyler}\ \emph {et~al.}(2019)\citenamefont {Seyler},
  \citenamefont {Rivera}, \citenamefont {Yu}, \citenamefont {Wilson},
  \citenamefont {Ray}, \citenamefont {Mandrus}, \citenamefont {Yan},
  \citenamefont {Yao},\ and\ \citenamefont {Xu}}]{seyler2019signatures}%
  \BibitemOpen
  \bibfield  {author} {\bibinfo {author} {\bibfnamefont {K.~L.}\ \bibnamefont
  {Seyler}}, \bibinfo {author} {\bibfnamefont {P.}~\bibnamefont {Rivera}},
  \bibinfo {author} {\bibfnamefont {H.}~\bibnamefont {Yu}}, \bibinfo {author}
  {\bibfnamefont {N.~P.}\ \bibnamefont {Wilson}}, \bibinfo {author}
  {\bibfnamefont {E.~L.}\ \bibnamefont {Ray}}, \bibinfo {author} {\bibfnamefont
  {D.~G.}\ \bibnamefont {Mandrus}}, \bibinfo {author} {\bibfnamefont
  {J.}~\bibnamefont {Yan}}, \bibinfo {author} {\bibfnamefont {W.}~\bibnamefont
  {Yao}}, \ and\ \bibinfo {author} {\bibfnamefont {X.}~\bibnamefont {Xu}},\
  }\bibfield  {title} {\enquote {\bibinfo {title} {Signatures of
  moir{\'e}-trapped valley excitons in MoSe$_2$/WSe$_2$ heterobilayers},}\
  }\href@noop {} {\bibfield  {journal} {\bibinfo  {journal} {Nature}\ }\textbf
  {\bibinfo {volume} {567}},\ \bibinfo {pages} {66} (\bibinfo {year}
  {2019})}\BibitemShut {NoStop}%
\bibitem [{\citenamefont {Tran}\ \emph {et~al.}(2019)\citenamefont {Tran},
  \citenamefont {Moody}, \citenamefont {Wu}, \citenamefont {Lu}, \citenamefont
  {Choi}, \citenamefont {Kim}, \citenamefont {Rai}, \citenamefont {Sanchez},
  \citenamefont {Quan}, \citenamefont {Singh}, \citenamefont {Embley},
  \citenamefont {Zepeda}, \citenamefont {Campbell}, \citenamefont {Autry},
  \citenamefont {Taniguchi}, \citenamefont {Watanabe}, \citenamefont {Lu},
  \citenamefont {Banerjee}, \citenamefont {Silverman}, \citenamefont {Kim},
  \citenamefont {Tutuc}, \citenamefont {Yang}, \citenamefont {MacDonald},\ and\
  \citenamefont {Li}}]{tran2019evidence}%
  \BibitemOpen
  \bibfield  {author} {\bibinfo {author} {\bibfnamefont {K.}~\bibnamefont
  {Tran}}, \bibinfo {author} {\bibfnamefont {G.}~\bibnamefont {Moody}},
  \bibinfo {author} {\bibfnamefont {F.}~\bibnamefont {Wu}}, \bibinfo {author}
  {\bibfnamefont {X.}~\bibnamefont {Lu}}, \bibinfo {author} {\bibfnamefont
  {J.}~\bibnamefont {Choi}}, \bibinfo {author} {\bibfnamefont {K.}~\bibnamefont
  {Kim}}, \bibinfo {author} {\bibfnamefont {A.}~\bibnamefont {Rai}}, \bibinfo
  {author} {\bibfnamefont {D.~A.}\ \bibnamefont {Sanchez}}, \bibinfo {author}
  {\bibfnamefont {J.}~\bibnamefont {Quan}}, \bibinfo {author} {\bibfnamefont
  {A.}~\bibnamefont {Singh}}, \bibinfo {author} {\bibfnamefont
  {J.}~\bibnamefont {Embley}}, \bibinfo {author} {\bibfnamefont
  {A.}~\bibnamefont {Zepeda}}, \bibinfo {author} {\bibfnamefont
  {M.}~\bibnamefont {Campbell}}, \bibinfo {author} {\bibfnamefont
  {T.}~\bibnamefont {Autry}}, \bibinfo {author} {\bibfnamefont
  {T.}~\bibnamefont {Taniguchi}}, \bibinfo {author} {\bibfnamefont
  {K.}~\bibnamefont {Watanabe}}, \bibinfo {author} {\bibfnamefont
  {N.}~\bibnamefont {Lu}}, \bibinfo {author} {\bibfnamefont {S.~K.}\
  \bibnamefont {Banerjee}}, \bibinfo {author} {\bibfnamefont {K.~L.}\
  \bibnamefont {Silverman}}, \bibinfo {author} {\bibfnamefont {S.}~\bibnamefont
  {Kim}}, \bibinfo {author} {\bibfnamefont {E.}~\bibnamefont {Tutuc}}, \bibinfo
  {author} {\bibfnamefont {L.}~\bibnamefont {Yang}}, \bibinfo {author}
  {\bibfnamefont {A.~H.}\ \bibnamefont {MacDonald}}, \ and\ \bibinfo {author}
  {\bibfnamefont {X.}~\bibnamefont {Li}},\ }\bibfield  {title} {\enquote
  {\bibinfo {title} {Evidence for moir{\'e} excitons in van der Waals
  heterostructures},}\ }\href {https://doi.org/10.1038/s41586-019-0975-z}
  {\bibfield  {journal} {\bibinfo  {journal} {Nature}\ }\textbf {\bibinfo
  {volume} {567}},\ \bibinfo {pages} {71} (\bibinfo {year} {2019})}\BibitemShut
  {NoStop}%
\bibitem [{\citenamefont {Jin}\ \emph {et~al.}(2019{\natexlab{a}})\citenamefont
  {Jin}, \citenamefont {Regan}, \citenamefont {Yan}, \citenamefont {Iqbal
  Bakti~Utama}, \citenamefont {Wang}, \citenamefont {Zhao}, \citenamefont
  {Qin}, \citenamefont {Yang}, \citenamefont {Zheng}, \citenamefont {Shi},
  \citenamefont {Watanabe}, \citenamefont {Taniguchi}, \citenamefont {Tongay},
  \citenamefont {Zettl},\ and\ \citenamefont {Wang}}]{jin2019observation}%
  \BibitemOpen
  \bibfield  {author} {\bibinfo {author} {\bibfnamefont {C.}~\bibnamefont
  {Jin}}, \bibinfo {author} {\bibfnamefont {E.~C.}\ \bibnamefont {Regan}},
  \bibinfo {author} {\bibfnamefont {A.}~\bibnamefont {Yan}}, \bibinfo {author}
  {\bibfnamefont {M.}~\bibnamefont {Iqbal Bakti~Utama}}, \bibinfo {author}
  {\bibfnamefont {D.}~\bibnamefont {Wang}}, \bibinfo {author} {\bibfnamefont
  {S.}~\bibnamefont {Zhao}}, \bibinfo {author} {\bibfnamefont {Y.}~\bibnamefont
  {Qin}}, \bibinfo {author} {\bibfnamefont {S.}~\bibnamefont {Yang}}, \bibinfo
  {author} {\bibfnamefont {Z.}~\bibnamefont {Zheng}}, \bibinfo {author}
  {\bibfnamefont {S.}~\bibnamefont {Shi}}, \bibinfo {author} {\bibfnamefont
  {K.}~\bibnamefont {Watanabe}}, \bibinfo {author} {\bibfnamefont
  {T.}~\bibnamefont {Taniguchi}}, \bibinfo {author} {\bibfnamefont
  {S.}~\bibnamefont {Tongay}}, \bibinfo {author} {\bibfnamefont
  {A.}~\bibnamefont {Zettl}}, \ and\ \bibinfo {author} {\bibfnamefont
  {F.}~\bibnamefont {Wang}},\ }\bibfield  {title} {\enquote {\bibinfo {title}
  {Observation of moir{\'e} excitons in WSe$_2$/WS$_2$ heterostructure
  superlattices},}\ }\href {https://doi.org/10.1038/s41586-019-0976-y}
  {\bibfield  {journal} {\bibinfo  {journal} {Nature}\ }\textbf {\bibinfo
  {volume} {567}},\ \bibinfo {pages} {76} (\bibinfo {year}
  {2019}{\natexlab{a}})}\BibitemShut {NoStop}%
\bibitem [{\citenamefont {Alexeev}\ \emph {et~al.}(2019)\citenamefont
  {Alexeev}, \citenamefont {Ruiz-Tijerina}, \citenamefont {Danovich},
  \citenamefont {Hamer}, \citenamefont {Terry}, \citenamefont {Nayak},
  \citenamefont {Ahn}, \citenamefont {Pak}, \citenamefont {Lee}, \citenamefont
  {Sohn}, \citenamefont {Molas}, \citenamefont {Koperski}, \citenamefont
  {Watanabe}, \citenamefont {Taniguchi}, \citenamefont {Novoselov},
  \citenamefont {Gorbachev}, \citenamefont {Shin}, \citenamefont {Fal'ko},\
  and\ \citenamefont {Tartakovskii}}]{alexeev2019resonantly}%
  \BibitemOpen
  \bibfield  {author} {\bibinfo {author} {\bibfnamefont {E.~M.}\ \bibnamefont
  {Alexeev}}, \bibinfo {author} {\bibfnamefont {D.~A.}\ \bibnamefont
  {Ruiz-Tijerina}}, \bibinfo {author} {\bibfnamefont {M.}~\bibnamefont
  {Danovich}}, \bibinfo {author} {\bibfnamefont {M.~J.}\ \bibnamefont {Hamer}},
  \bibinfo {author} {\bibfnamefont {D.~J.}\ \bibnamefont {Terry}}, \bibinfo
  {author} {\bibfnamefont {P.~K.}\ \bibnamefont {Nayak}}, \bibinfo {author}
  {\bibfnamefont {S.}~\bibnamefont {Ahn}}, \bibinfo {author} {\bibfnamefont
  {S.}~\bibnamefont {Pak}}, \bibinfo {author} {\bibfnamefont {J.}~\bibnamefont
  {Lee}}, \bibinfo {author} {\bibfnamefont {J.~I.}\ \bibnamefont {Sohn}},
  \bibinfo {author} {\bibfnamefont {M.~R.}\ \bibnamefont {Molas}}, \bibinfo
  {author} {\bibfnamefont {M.}~\bibnamefont {Koperski}}, \bibinfo {author}
  {\bibfnamefont {K.}~\bibnamefont {Watanabe}}, \bibinfo {author}
  {\bibfnamefont {T.}~\bibnamefont {Taniguchi}}, \bibinfo {author}
  {\bibfnamefont {K.~S.}\ \bibnamefont {Novoselov}}, \bibinfo {author}
  {\bibfnamefont {R.~V.}\ \bibnamefont {Gorbachev}}, \bibinfo {author}
  {\bibfnamefont {H.~S.}\ \bibnamefont {Shin}}, \bibinfo {author}
  {\bibfnamefont {V.~I.}\ \bibnamefont {Fal'ko}}, \ and\ \bibinfo {author}
  {\bibfnamefont {A.~I.}\ \bibnamefont {Tartakovskii}},\ }\bibfield  {title}
  {\enquote {\bibinfo {title} {Resonantly hybridized excitons in moir{\'e}
  superlattices in van der Waals heterostructures},}\ }\href
  {https://doi.org/10.1038/s41586-019-0986-9} {\bibfield  {journal} {\bibinfo
  {journal} {Nature}\ }\textbf {\bibinfo {volume} {567}},\ \bibinfo {pages}
  {81} (\bibinfo {year} {2019})}\BibitemShut {NoStop}%
\bibitem [{\citenamefont {Tang}\ \emph {et~al.}(2020)\citenamefont {Tang},
  \citenamefont {Li}, \citenamefont {Li}, \citenamefont {Xu}, \citenamefont
  {Liu}, \citenamefont {Barmak}, \citenamefont {Watanabe}, \citenamefont
  {Taniguchi}, \citenamefont {MacDonald}, \citenamefont {Shan},\ and\
  \citenamefont {Mak}}]{tang2020simulation}%
  \BibitemOpen
  \bibfield  {author} {\bibinfo {author} {\bibfnamefont {Y.}~\bibnamefont
  {Tang}}, \bibinfo {author} {\bibfnamefont {L.}~\bibnamefont {Li}}, \bibinfo
  {author} {\bibfnamefont {T.}~\bibnamefont {Li}}, \bibinfo {author}
  {\bibfnamefont {Y.}~\bibnamefont {Xu}}, \bibinfo {author} {\bibfnamefont
  {S.}~\bibnamefont {Liu}}, \bibinfo {author} {\bibfnamefont {K.}~\bibnamefont
  {Barmak}}, \bibinfo {author} {\bibfnamefont {K.}~\bibnamefont {Watanabe}},
  \bibinfo {author} {\bibfnamefont {T.}~\bibnamefont {Taniguchi}}, \bibinfo
  {author} {\bibfnamefont {A.~H.}\ \bibnamefont {MacDonald}}, \bibinfo {author}
  {\bibfnamefont {J.}~\bibnamefont {Shan}}, \ and\ \bibinfo {author}
  {\bibfnamefont {K.~F.}\ \bibnamefont {Mak}},\ }\bibfield  {title} {\enquote
  {\bibinfo {title} {Simulation of Hubbard model physics in WSe2/WS2 moir{\'e}
  superlattices},}\ }\href {https://doi.org/10.1038/s41586-020-2085-3}
  {\bibfield  {journal} {\bibinfo  {journal} {Nature}\ }\textbf {\bibinfo
  {volume} {579}},\ \bibinfo {pages} {353} (\bibinfo {year}
  {2020})}\BibitemShut {NoStop}%
\bibitem [{\citenamefont {Regan}\ \emph {et~al.}(2020)\citenamefont {Regan},
  \citenamefont {Wang}, \citenamefont {Jin}, \citenamefont {Bakti~Utama},
  \citenamefont {Gao}, \citenamefont {Wei}, \citenamefont {Zhao}, \citenamefont
  {Zhao}, \citenamefont {Zhang}, \citenamefont {Yumigeta}, \citenamefont
  {Blei}, \citenamefont {Carlstr{\"o}m}, \citenamefont {Watanabe},
  \citenamefont {Taniguchi}, \citenamefont {Tongay}, \citenamefont {Crommie},
  \citenamefont {Zettl},\ and\ \citenamefont {Wang}}]{regan2020mott}%
  \BibitemOpen
  \bibfield  {author} {\bibinfo {author} {\bibfnamefont {E.~C.}\ \bibnamefont
  {Regan}}, \bibinfo {author} {\bibfnamefont {D.}~\bibnamefont {Wang}},
  \bibinfo {author} {\bibfnamefont {C.}~\bibnamefont {Jin}}, \bibinfo {author}
  {\bibfnamefont {M.~I.}\ \bibnamefont {Bakti~Utama}}, \bibinfo {author}
  {\bibfnamefont {B.}~\bibnamefont {Gao}}, \bibinfo {author} {\bibfnamefont
  {X.}~\bibnamefont {Wei}}, \bibinfo {author} {\bibfnamefont {S.}~\bibnamefont
  {Zhao}}, \bibinfo {author} {\bibfnamefont {W.}~\bibnamefont {Zhao}}, \bibinfo
  {author} {\bibfnamefont {Z.}~\bibnamefont {Zhang}}, \bibinfo {author}
  {\bibfnamefont {K.}~\bibnamefont {Yumigeta}}, \bibinfo {author}
  {\bibfnamefont {M.}~\bibnamefont {Blei}}, \bibinfo {author} {\bibfnamefont
  {J.~D.}\ \bibnamefont {Carlstr{\"o}m}}, \bibinfo {author} {\bibfnamefont
  {K.}~\bibnamefont {Watanabe}}, \bibinfo {author} {\bibfnamefont
  {T.}~\bibnamefont {Taniguchi}}, \bibinfo {author} {\bibfnamefont
  {S.}~\bibnamefont {Tongay}}, \bibinfo {author} {\bibfnamefont
  {M.}~\bibnamefont {Crommie}}, \bibinfo {author} {\bibfnamefont
  {A.}~\bibnamefont {Zettl}}, \ and\ \bibinfo {author} {\bibfnamefont
  {F.}~\bibnamefont {Wang}},\ }\bibfield  {title} {\enquote {\bibinfo {title}
  {Mott and generalized Wigner crystal states in WSe$_2$/WS$_2$ moir{\'e}
  superlattices},}\ }\href {https://doi.org/10.1038/s41586-020-2092-4}
  {\bibfield  {journal} {\bibinfo  {journal} {Nature}\ }\textbf {\bibinfo
  {volume} {579}},\ \bibinfo {pages} {359} (\bibinfo {year}
  {2020})}\BibitemShut {NoStop}%
\bibitem [{\citenamefont {Shimazaki}\ \emph {et~al.}(2020)\citenamefont
  {Shimazaki}, \citenamefont {Schwartz}, \citenamefont {Watanabe},
  \citenamefont {Taniguchi}, \citenamefont {Kroner},\ and\ \citenamefont
  {Imamo{\u{g}}lu}}]{shimazaki2020strongly}%
  \BibitemOpen
  \bibfield  {author} {\bibinfo {author} {\bibfnamefont {Y.}~\bibnamefont
  {Shimazaki}}, \bibinfo {author} {\bibfnamefont {I.}~\bibnamefont {Schwartz}},
  \bibinfo {author} {\bibfnamefont {K.}~\bibnamefont {Watanabe}}, \bibinfo
  {author} {\bibfnamefont {T.}~\bibnamefont {Taniguchi}}, \bibinfo {author}
  {\bibfnamefont {M.}~\bibnamefont {Kroner}}, \ and\ \bibinfo {author}
  {\bibfnamefont {A.}~\bibnamefont {Imamo{\u{g}}lu}},\ }\bibfield  {title}
  {\enquote {\bibinfo {title} {Strongly correlated electrons and hybrid
  excitons in a moiré heterostructure},}\ }\href
  {https://doi.org/10.1038/s41586-020-2191-2} {\bibfield  {journal} {\bibinfo
  {journal} {Nature}\ }\textbf {\bibinfo {volume} {580}},\ \bibinfo {pages}
  {472} (\bibinfo {year} {2020})}\BibitemShut {NoStop}%
\bibitem [{\citenamefont {Wang}\ \emph {et~al.}(2020)\citenamefont {Wang},
  \citenamefont {Shih}, \citenamefont {Ghiotto}, \citenamefont {Xian},
  \citenamefont {Rhodes}, \citenamefont {Tan}, \citenamefont {Claassen},
  \citenamefont {Kennes}, \citenamefont {Bai}, \citenamefont {Kim},
  \citenamefont {Watanabe}, \citenamefont {Taniguchi}, \citenamefont {Zhu},
  \citenamefont {Hone}, \citenamefont {Rubio}, \citenamefont {Pasupathy},\ and\
  \citenamefont {Dean}}]{wang2020correlated}%
  \BibitemOpen
  \bibfield  {author} {\bibinfo {author} {\bibfnamefont {L.}~\bibnamefont
  {Wang}}, \bibinfo {author} {\bibfnamefont {E.-M.}\ \bibnamefont {Shih}},
  \bibinfo {author} {\bibfnamefont {A.}~\bibnamefont {Ghiotto}}, \bibinfo
  {author} {\bibfnamefont {L.}~\bibnamefont {Xian}}, \bibinfo {author}
  {\bibfnamefont {D.~A.}\ \bibnamefont {Rhodes}}, \bibinfo {author}
  {\bibfnamefont {C.}~\bibnamefont {Tan}}, \bibinfo {author} {\bibfnamefont
  {M.}~\bibnamefont {Claassen}}, \bibinfo {author} {\bibfnamefont {D.~M.}\
  \bibnamefont {Kennes}}, \bibinfo {author} {\bibfnamefont {Y.}~\bibnamefont
  {Bai}}, \bibinfo {author} {\bibfnamefont {B.}~\bibnamefont {Kim}}, \bibinfo
  {author} {\bibfnamefont {K.}~\bibnamefont {Watanabe}}, \bibinfo {author}
  {\bibfnamefont {T.}~\bibnamefont {Taniguchi}}, \bibinfo {author}
  {\bibfnamefont {X.}~\bibnamefont {Zhu}}, \bibinfo {author} {\bibfnamefont
  {J.}~\bibnamefont {Hone}}, \bibinfo {author} {\bibfnamefont {A.}~\bibnamefont
  {Rubio}}, \bibinfo {author} {\bibfnamefont {A.~N.}\ \bibnamefont
  {Pasupathy}}, \ and\ \bibinfo {author} {\bibfnamefont {C.~R.}\ \bibnamefont
  {Dean}},\ }\bibfield  {title} {\enquote {\bibinfo {title} {Correlated
  electronic phases in twisted bilayer transition metal dichalcogenides},}\
  }\href {https://doi.org/10.1038/s41563-020-0708-6} {\bibfield  {journal}
  {\bibinfo  {journal} {Nat. Mater.}\ }\textbf {\bibinfo {volume} {19}},\
  \bibinfo {pages} {861} (\bibinfo {year} {2020})}\BibitemShut {NoStop}%
\bibitem [{\citenamefont {Weston}\ \emph {et~al.}(2020)\citenamefont {Weston},
  \citenamefont {Zou}, \citenamefont {Enaldiev}, \citenamefont {Summerfield},
  \citenamefont {Clark}, \citenamefont {Z{\'o}lyomi}, \citenamefont {Graham},
  \citenamefont {Yelgel}, \citenamefont {Magorrian}, \citenamefont {Zhou},
  \citenamefont {Zultak}, \citenamefont {Hopkinson}, \citenamefont {Barinov},
  \citenamefont {Bointon}, \citenamefont {Kretinin}, \citenamefont {Wilson},
  \citenamefont {Beton}, \citenamefont {Fal'ko}, \citenamefont {Haigh},\ and\
  \citenamefont {Gorbachev}}]{weston2020atomic}%
  \BibitemOpen
  \bibfield  {author} {\bibinfo {author} {\bibfnamefont {A.}~\bibnamefont
  {Weston}}, \bibinfo {author} {\bibfnamefont {Y.}~\bibnamefont {Zou}},
  \bibinfo {author} {\bibfnamefont {V.}~\bibnamefont {Enaldiev}}, \bibinfo
  {author} {\bibfnamefont {A.}~\bibnamefont {Summerfield}}, \bibinfo {author}
  {\bibfnamefont {N.}~\bibnamefont {Clark}}, \bibinfo {author} {\bibfnamefont
  {V.}~\bibnamefont {Z{\'o}lyomi}}, \bibinfo {author} {\bibfnamefont
  {A.}~\bibnamefont {Graham}}, \bibinfo {author} {\bibfnamefont
  {C.}~\bibnamefont {Yelgel}}, \bibinfo {author} {\bibfnamefont
  {S.}~\bibnamefont {Magorrian}}, \bibinfo {author} {\bibfnamefont
  {M.}~\bibnamefont {Zhou}}, \bibinfo {author} {\bibfnamefont {J.}~\bibnamefont
  {Zultak}}, \bibinfo {author} {\bibfnamefont {D.}~\bibnamefont {Hopkinson}},
  \bibinfo {author} {\bibfnamefont {A.}~\bibnamefont {Barinov}}, \bibinfo
  {author} {\bibfnamefont {T.~H.}\ \bibnamefont {Bointon}}, \bibinfo {author}
  {\bibfnamefont {A.}~\bibnamefont {Kretinin}}, \bibinfo {author}
  {\bibfnamefont {N.~R.}\ \bibnamefont {Wilson}}, \bibinfo {author}
  {\bibfnamefont {P.~H.}\ \bibnamefont {Beton}}, \bibinfo {author}
  {\bibfnamefont {V.~I.}\ \bibnamefont {Fal'ko}}, \bibinfo {author}
  {\bibfnamefont {S.~J.}\ \bibnamefont {Haigh}}, \ and\ \bibinfo {author}
  {\bibfnamefont {R.}~\bibnamefont {Gorbachev}},\ }\bibfield  {title} {\enquote
  {\bibinfo {title} {Atomic reconstruction in twisted bilayers of transition
  metal dichalcogenides},}\ }\href {https://doi.org/10.1038/s41565-020-0682-9}
  {\bibfield  {journal} {\bibinfo  {journal} {Nat. Nanotechnol.}\ }\textbf
  {\bibinfo {volume} {15}},\ \bibinfo {pages} {592} (\bibinfo {year}
  {2020})}\BibitemShut {NoStop}%
\bibitem [{\citenamefont {Rosenberger}\ \emph {et~al.}(2020)\citenamefont
  {Rosenberger}, \citenamefont {Chuang}, \citenamefont {Phillips},
  \citenamefont {Oleshko}, \citenamefont {McCreary}, \citenamefont {Sivaram},
  \citenamefont {Hellberg},\ and\ \citenamefont
  {Jonker}}]{doi:10.1021/acsnano.0c04832}%
  \BibitemOpen
  \bibfield  {author} {\bibinfo {author} {\bibfnamefont {M.~R.}\ \bibnamefont
  {Rosenberger}}, \bibinfo {author} {\bibfnamefont {H.-J.}\ \bibnamefont
  {Chuang}}, \bibinfo {author} {\bibfnamefont {M.}~\bibnamefont {Phillips}},
  \bibinfo {author} {\bibfnamefont {V.~P.}\ \bibnamefont {Oleshko}}, \bibinfo
  {author} {\bibfnamefont {K.~M.}\ \bibnamefont {McCreary}}, \bibinfo {author}
  {\bibfnamefont {S.~V.}\ \bibnamefont {Sivaram}}, \bibinfo {author}
  {\bibfnamefont {C.~S.}\ \bibnamefont {Hellberg}}, \ and\ \bibinfo {author}
  {\bibfnamefont {B.~T.}\ \bibnamefont {Jonker}},\ }\bibfield  {title}
  {\enquote {\bibinfo {title} {Correction to Twist Angle-Dependent Atomic
  Reconstruction and Moir{\'e} Patterns in Transition Metal Dichalcogenide
  Heterostructures},}\ }\href {\doibase 10.1021/acsnano.0c04832} {\bibfield
  {journal} {\bibinfo  {journal} {ACS Nano}\ }\textbf {\bibinfo {volume}
  {14}},\ \bibinfo {pages} {14240} (\bibinfo {year} {2020})}\BibitemShut
  {NoStop}%
\bibitem [{\citenamefont {Yu}\ \emph {et~al.}(2017)\citenamefont {Yu},
  \citenamefont {Liu}, \citenamefont {Tang}, \citenamefont {Xu},\ and\
  \citenamefont {Yao}}]{Yue1701696}%
  \BibitemOpen
  \bibfield  {author} {\bibinfo {author} {\bibfnamefont {H.}~\bibnamefont
  {Yu}}, \bibinfo {author} {\bibfnamefont {G.-B.}\ \bibnamefont {Liu}},
  \bibinfo {author} {\bibfnamefont {J.}~\bibnamefont {Tang}}, \bibinfo {author}
  {\bibfnamefont {X.}~\bibnamefont {Xu}}, \ and\ \bibinfo {author}
  {\bibfnamefont {W.}~\bibnamefont {Yao}},\ }\bibfield  {title} {\enquote
  {\bibinfo {title} {Moir{\'e} excitons: From programmable quantum emitter
  arrays to spin-orbit{\textendash}coupled artificial lattices},}\ }\href
  {\doibase 10.1126/sciadv.1701696} {\bibfield  {journal} {\bibinfo  {journal}
  {Sci. Adv.}\ }\textbf {\bibinfo {volume} {3}} (\bibinfo {year}
  {2017})}\BibitemShut {NoStop}%
\bibitem [{\citenamefont {Brotons-Gisbert}\ \emph {et~al.}(2020)\citenamefont
  {Brotons-Gisbert}, \citenamefont {Baek}, \citenamefont {Molina-S{\'a}nchez},
  \citenamefont {Campbell}, \citenamefont {Scerri}, \citenamefont {White},
  \citenamefont {Watanabe}, \citenamefont {Taniguchi}, \citenamefont {Bonato},\
  and\ \citenamefont {Gerardot}}]{brotons2020spin}%
  \BibitemOpen
  \bibfield  {author} {\bibinfo {author} {\bibfnamefont {M.}~\bibnamefont
  {Brotons-Gisbert}}, \bibinfo {author} {\bibfnamefont {H.}~\bibnamefont
  {Baek}}, \bibinfo {author} {\bibfnamefont {A.}~\bibnamefont
  {Molina-S{\'a}nchez}}, \bibinfo {author} {\bibfnamefont {A.}~\bibnamefont
  {Campbell}}, \bibinfo {author} {\bibfnamefont {E.}~\bibnamefont {Scerri}},
  \bibinfo {author} {\bibfnamefont {D.}~\bibnamefont {White}}, \bibinfo
  {author} {\bibfnamefont {K.}~\bibnamefont {Watanabe}}, \bibinfo {author}
  {\bibfnamefont {T.}~\bibnamefont {Taniguchi}}, \bibinfo {author}
  {\bibfnamefont {C.}~\bibnamefont {Bonato}}, \ and\ \bibinfo {author}
  {\bibfnamefont {B.~D.}\ \bibnamefont {Gerardot}},\ }\bibfield  {title}
  {\enquote {\bibinfo {title} {Spin--layer locking of interlayer excitons
  trapped in moir{\'e} potentials},}\ }\href
  {https://doi.org/10.1038/s41563-020-0687-7} {\bibfield  {journal} {\bibinfo
  {journal} {Nat. Mater.}\ }\textbf {\bibinfo {volume} {19}},\ \bibinfo {pages}
  {630} (\bibinfo {year} {2020})}\BibitemShut {NoStop}%
\bibitem [{\citenamefont {Shabani}\ \emph {et~al.}(2021)\citenamefont
  {Shabani}, \citenamefont {Halbertal}, \citenamefont {Wu}, \citenamefont
  {Chen}, \citenamefont {Liu}, \citenamefont {Hone}, \citenamefont {Yao},
  \citenamefont {Basov}, \citenamefont {Zhu},\ and\ \citenamefont
  {Pasupathy}}]{shabani2021deep}%
  \BibitemOpen
  \bibfield  {author} {\bibinfo {author} {\bibfnamefont {S.}~\bibnamefont
  {Shabani}}, \bibinfo {author} {\bibfnamefont {D.}~\bibnamefont {Halbertal}},
  \bibinfo {author} {\bibfnamefont {W.}~\bibnamefont {Wu}}, \bibinfo {author}
  {\bibfnamefont {M.}~\bibnamefont {Chen}}, \bibinfo {author} {\bibfnamefont
  {S.}~\bibnamefont {Liu}}, \bibinfo {author} {\bibfnamefont {J.}~\bibnamefont
  {Hone}}, \bibinfo {author} {\bibfnamefont {W.}~\bibnamefont {Yao}}, \bibinfo
  {author} {\bibfnamefont {D.~N.}\ \bibnamefont {Basov}}, \bibinfo {author}
  {\bibfnamefont {X.}~\bibnamefont {Zhu}}, \ and\ \bibinfo {author}
  {\bibfnamefont {A.~N.}\ \bibnamefont {Pasupathy}},\ }\bibfield  {title}
  {\enquote {\bibinfo {title} {Deep moir{\'e} potentials in twisted transition
  metal dichalcogenide bilayers},}\ }\href
  {https://doi.org/10.1038/s41567-021-01174-7} {\bibfield  {journal} {\bibinfo
  {journal} {Nat. Phys.}\ }\textbf {\bibinfo {volume} {17}},\ \bibinfo {pages}
  {720} (\bibinfo {year} {2021})}\BibitemShut {NoStop}%
\bibitem [{\citenamefont {Jin}\ \emph {et~al.}(2019{\natexlab{b}})\citenamefont
  {Jin}, \citenamefont {Regan}, \citenamefont {Wang}, \citenamefont {Iqbal
  Bakti~Utama}, \citenamefont {Yang}, \citenamefont {Cain}, \citenamefont
  {Qin}, \citenamefont {Shen}, \citenamefont {Zheng}, \citenamefont {Watanabe},
  \citenamefont {Taniguchi}, \citenamefont {Tongay}, \citenamefont {Zettl},\
  and\ \citenamefont {Wang}}]{jin2019identification}%
  \BibitemOpen
  \bibfield  {author} {\bibinfo {author} {\bibfnamefont {C.}~\bibnamefont
  {Jin}}, \bibinfo {author} {\bibfnamefont {E.~C.}\ \bibnamefont {Regan}},
  \bibinfo {author} {\bibfnamefont {D.}~\bibnamefont {Wang}}, \bibinfo {author}
  {\bibfnamefont {M.}~\bibnamefont {Iqbal Bakti~Utama}}, \bibinfo {author}
  {\bibfnamefont {C.-S.}\ \bibnamefont {Yang}}, \bibinfo {author}
  {\bibfnamefont {J.}~\bibnamefont {Cain}}, \bibinfo {author} {\bibfnamefont
  {Y.}~\bibnamefont {Qin}}, \bibinfo {author} {\bibfnamefont {Y.}~\bibnamefont
  {Shen}}, \bibinfo {author} {\bibfnamefont {Z.}~\bibnamefont {Zheng}},
  \bibinfo {author} {\bibfnamefont {K.}~\bibnamefont {Watanabe}}, \bibinfo
  {author} {\bibfnamefont {T.}~\bibnamefont {Taniguchi}}, \bibinfo {author}
  {\bibfnamefont {S.}~\bibnamefont {Tongay}}, \bibinfo {author} {\bibfnamefont
  {A.}~\bibnamefont {Zettl}}, \ and\ \bibinfo {author} {\bibfnamefont
  {F.}~\bibnamefont {Wang}},\ }\bibfield  {title} {\enquote {\bibinfo {title}
  {Identification of spin, valley and moir{\'e} quasi-angular momentum of
  interlayer excitons},}\ }\href {https://doi.org/10.1038/s41567-019-0631-4}
  {\bibfield  {journal} {\bibinfo  {journal} {Nat. Phys.}\ }\textbf {\bibinfo
  {volume} {15}},\ \bibinfo {pages} {1140} (\bibinfo {year}
  {2019}{\natexlab{b}})}\BibitemShut {NoStop}%
\bibitem [{\citenamefont {Baek}\ \emph {et~al.}(2020)\citenamefont {Baek},
  \citenamefont {Brotons-Gisbert}, \citenamefont {Koong}, \citenamefont
  {Campbell}, \citenamefont {Rambach}, \citenamefont {Watanabe}, \citenamefont
  {Taniguchi},\ and\ \citenamefont {Gerardot}}]{Baekeaba8526}%
  \BibitemOpen
  \bibfield  {author} {\bibinfo {author} {\bibfnamefont {H.}~\bibnamefont
  {Baek}}, \bibinfo {author} {\bibfnamefont {M.}~\bibnamefont
  {Brotons-Gisbert}}, \bibinfo {author} {\bibfnamefont {Z.~X.}\ \bibnamefont
  {Koong}}, \bibinfo {author} {\bibfnamefont {A.}~\bibnamefont {Campbell}},
  \bibinfo {author} {\bibfnamefont {M.}~\bibnamefont {Rambach}}, \bibinfo
  {author} {\bibfnamefont {K.}~\bibnamefont {Watanabe}}, \bibinfo {author}
  {\bibfnamefont {T.}~\bibnamefont {Taniguchi}}, \ and\ \bibinfo {author}
  {\bibfnamefont {B.~D.}\ \bibnamefont {Gerardot}},\ }\bibfield  {title}
  {\enquote {\bibinfo {title} {Highly energy-tunable quantum light from
  moir{\'e}-trapped excitons},}\ }\href {\doibase 10.1126/sciadv.aba8526}
  {\bibfield  {journal} {\bibinfo  {journal} {Sci. Adv.}\ }\textbf {\bibinfo
  {volume} {6}} (\bibinfo {year} {2020})}\BibitemShut {NoStop}%
\bibitem [{\citenamefont {Liu}\ \emph {et~al.}(2021)\citenamefont {Liu},
  \citenamefont {Barr{\'e}}, \citenamefont {van Baren}, \citenamefont {Wilson},
  \citenamefont {Taniguchi}, \citenamefont {Watanabe}, \citenamefont {Cui},
  \citenamefont {Gabor}, \citenamefont {Heinz}, \citenamefont {Chang},\ and\
  \citenamefont {Lui}}]{liu2021signatures}%
  \BibitemOpen
  \bibfield  {author} {\bibinfo {author} {\bibfnamefont {E.}~\bibnamefont
  {Liu}}, \bibinfo {author} {\bibfnamefont {E.}~\bibnamefont {Barr{\'e}}},
  \bibinfo {author} {\bibfnamefont {J.}~\bibnamefont {van Baren}}, \bibinfo
  {author} {\bibfnamefont {M.}~\bibnamefont {Wilson}}, \bibinfo {author}
  {\bibfnamefont {T.}~\bibnamefont {Taniguchi}}, \bibinfo {author}
  {\bibfnamefont {K.}~\bibnamefont {Watanabe}}, \bibinfo {author}
  {\bibfnamefont {Y.-T.}\ \bibnamefont {Cui}}, \bibinfo {author} {\bibfnamefont
  {N.~M.}\ \bibnamefont {Gabor}}, \bibinfo {author} {\bibfnamefont {T.~F.}\
  \bibnamefont {Heinz}}, \bibinfo {author} {\bibfnamefont {Y.-C.}\ \bibnamefont
  {Chang}}, \ and\ \bibinfo {author} {\bibfnamefont {C.~H.}\ \bibnamefont
  {Lui}},\ }\bibfield  {title} {\enquote {\bibinfo {title} {Signatures of
  moir{\'e} trions in WSe$_2$/MoSe$_2$ heterobilayers},}\ }\href
  {https://doi.org/10.1038/s41586-021-03541-z} {\bibfield  {journal} {\bibinfo
  {journal} {Nature}\ }\textbf {\bibinfo {volume} {594}},\ \bibinfo {pages}
  {46} (\bibinfo {year} {2021})}\BibitemShut {NoStop}%
\bibitem [{\citenamefont {Brotons-Gisbert}\ \emph {et~al.}(2021)\citenamefont
  {Brotons-Gisbert}, \citenamefont {Baek}, \citenamefont {Campbell},
  \citenamefont {Watanabe}, \citenamefont {Taniguchi},\ and\ \citenamefont
  {Gerardot}}]{brotonsgisbert2021moiretrapped}%
  \BibitemOpen
  \bibfield  {author} {\bibinfo {author} {\bibfnamefont {M.}~\bibnamefont
  {Brotons-Gisbert}}, \bibinfo {author} {\bibfnamefont {H.}~\bibnamefont
  {Baek}}, \bibinfo {author} {\bibfnamefont {A.}~\bibnamefont {Campbell}},
  \bibinfo {author} {\bibfnamefont {K.}~\bibnamefont {Watanabe}}, \bibinfo
  {author} {\bibfnamefont {T.}~\bibnamefont {Taniguchi}}, \ and\ \bibinfo
  {author} {\bibfnamefont {B.~D.}\ \bibnamefont {Gerardot}},\ }\bibfield
  {title} {\enquote {\bibinfo {title} {Moir\'e-Trapped Interlayer Trions in a
  Charge-Tunable ${\mathrm{WSe}}_{2}/{\mathrm{MoSe}}_{2}$ Heterobilayer},}\
  }\href {\doibase 10.1103/PhysRevX.11.031033} {\bibfield  {journal} {\bibinfo
  {journal} {Phys. Rev. X}\ }\textbf {\bibinfo {volume} {11}},\ \bibinfo
  {pages} {031033} (\bibinfo {year} {2021})}\BibitemShut {NoStop}%
\bibitem [{\citenamefont {Baek}\ \emph {et~al.}(2021)\citenamefont {Baek},
  \citenamefont {Brotons-Gisbert}, \citenamefont {Campbell}, \citenamefont
  {Vitale}, \citenamefont {Lischner}, \citenamefont {Watanabe}, \citenamefont
  {Taniguchi},\ and\ \citenamefont {Gerardot}}]{baek2021optical}%
  \BibitemOpen
  \bibfield  {author} {\bibinfo {author} {\bibfnamefont {H.}~\bibnamefont
  {Baek}}, \bibinfo {author} {\bibfnamefont {M.}~\bibnamefont
  {Brotons-Gisbert}}, \bibinfo {author} {\bibfnamefont {A.}~\bibnamefont
  {Campbell}}, \bibinfo {author} {\bibfnamefont {V.}~\bibnamefont {Vitale}},
  \bibinfo {author} {\bibfnamefont {J.}~\bibnamefont {Lischner}}, \bibinfo
  {author} {\bibfnamefont {K.}~\bibnamefont {Watanabe}}, \bibinfo {author}
  {\bibfnamefont {T.}~\bibnamefont {Taniguchi}}, \ and\ \bibinfo {author}
  {\bibfnamefont {B.~D.}\ \bibnamefont {Gerardot}},\ }\href@noop {} {\enquote
  {\bibinfo {title} {Optical read-out of Coulomb staircases in a moir\'e
  superlattice via trapped interlayer trions},}\ } (\bibinfo {year} {2021}),\
  \Eprint {http://arxiv.org/abs/2102.01358} {arXiv:2102.01358} \BibitemShut
  {NoStop}%
\bibitem [{\citenamefont {F{\"o}rg}\ \emph {et~al.}(2019)\citenamefont
  {F{\"o}rg}, \citenamefont {Colombier}, \citenamefont {Patel}, \citenamefont
  {Lindlau}, \citenamefont {Mohite}, \citenamefont {Yamaguchi}, \citenamefont
  {Glazov}, \citenamefont {Hunger},\ and\ \citenamefont
  {H{\"o}gele}}]{forg2019cavity}%
  \BibitemOpen
  \bibfield  {author} {\bibinfo {author} {\bibfnamefont {M.}~\bibnamefont
  {F{\"o}rg}}, \bibinfo {author} {\bibfnamefont {L.}~\bibnamefont {Colombier}},
  \bibinfo {author} {\bibfnamefont {R.~K.}\ \bibnamefont {Patel}}, \bibinfo
  {author} {\bibfnamefont {J.}~\bibnamefont {Lindlau}}, \bibinfo {author}
  {\bibfnamefont {A.~D.}\ \bibnamefont {Mohite}}, \bibinfo {author}
  {\bibfnamefont {H.}~\bibnamefont {Yamaguchi}}, \bibinfo {author}
  {\bibfnamefont {M.~M.}\ \bibnamefont {Glazov}}, \bibinfo {author}
  {\bibfnamefont {D.}~\bibnamefont {Hunger}}, \ and\ \bibinfo {author}
  {\bibfnamefont {A.}~\bibnamefont {H{\"o}gele}},\ }\bibfield  {title}
  {\enquote {\bibinfo {title} {Cavity-control of interlayer excitons in van der
  Waals heterostructures},}\ }\href@noop {} {\bibfield  {journal} {\bibinfo
  {journal} {Nat. comm.}\ }\textbf {\bibinfo {volume} {10}},\ \bibinfo {pages}
  {3697} (\bibinfo {year} {2019})}\BibitemShut {NoStop}%
\bibitem [{\citenamefont {Kioseoglou}\ \emph {et~al.}(2012)\citenamefont
  {Kioseoglou}, \citenamefont {Hanbicki}, \citenamefont {Currie}, \citenamefont
  {Friedman}, \citenamefont {Gunlycke},\ and\ \citenamefont
  {Jonker}}]{Kioseoglou2012}%
  \BibitemOpen
  \bibfield  {author} {\bibinfo {author} {\bibfnamefont {G.}~\bibnamefont
  {Kioseoglou}}, \bibinfo {author} {\bibfnamefont {A.~T.}\ \bibnamefont
  {Hanbicki}}, \bibinfo {author} {\bibfnamefont {M.}~\bibnamefont {Currie}},
  \bibinfo {author} {\bibfnamefont {A.~L.}\ \bibnamefont {Friedman}}, \bibinfo
  {author} {\bibfnamefont {D.}~\bibnamefont {Gunlycke}}, \ and\ \bibinfo
  {author} {\bibfnamefont {B.~T.}\ \bibnamefont {Jonker}},\ }\bibfield  {title}
  {\enquote {\bibinfo {title} {Valley polarization and intervalley scattering
  in monolayer MoS$_2$},}\ }\href {\doibase 10.1063/1.4768299} {\bibfield
  {journal} {\bibinfo  {journal} {Appl. Phys. Lett.}\ }\textbf {\bibinfo
  {volume} {101}},\ \bibinfo {pages} {221907} (\bibinfo {year}
  {2012})}\BibitemShut {NoStop}%
\bibitem [{\citenamefont {Glazov}\ \emph {et~al.}(2014)\citenamefont {Glazov},
  \citenamefont {Amand}, \citenamefont {Marie}, \citenamefont {Lagarde},
  \citenamefont {Bouet},\ and\ \citenamefont {Urbaszek}}]{PhysRevB.89.201302}%
  \BibitemOpen
  \bibfield  {author} {\bibinfo {author} {\bibfnamefont {M.~M.}\ \bibnamefont
  {Glazov}}, \bibinfo {author} {\bibfnamefont {T.}~\bibnamefont {Amand}},
  \bibinfo {author} {\bibfnamefont {X.}~\bibnamefont {Marie}}, \bibinfo
  {author} {\bibfnamefont {D.}~\bibnamefont {Lagarde}}, \bibinfo {author}
  {\bibfnamefont {L.}~\bibnamefont {Bouet}}, \ and\ \bibinfo {author}
  {\bibfnamefont {B.}~\bibnamefont {Urbaszek}},\ }\bibfield  {title} {\enquote
  {\bibinfo {title} {Exciton fine structure and spin decoherence in monolayers
  of transition metal dichalcogenides},}\ }\href {\doibase
  10.1103/PhysRevB.89.201302} {\bibfield  {journal} {\bibinfo  {journal} {Phys.
  Rev. B}\ }\textbf {\bibinfo {volume} {89}},\ \bibinfo {pages} {201302}
  (\bibinfo {year} {2014})}\BibitemShut {NoStop}%
\bibitem [{\citenamefont {Yu}\ and\ \citenamefont
  {Wu}(2014)}]{PhysRevB.89.205303}%
  \BibitemOpen
  \bibfield  {author} {\bibinfo {author} {\bibfnamefont {T.}~\bibnamefont
  {Yu}}\ and\ \bibinfo {author} {\bibfnamefont {M.~W.}\ \bibnamefont {Wu}},\
  }\bibfield  {title} {\enquote {\bibinfo {title} {Valley depolarization due to
  intervalley and intravalley electron-hole exchange interactions in monolayer
  ${\text{MoS}}_{2}$},}\ }\href {\doibase 10.1103/PhysRevB.89.205303}
  {\bibfield  {journal} {\bibinfo  {journal} {Phys. Rev. B}\ }\textbf {\bibinfo
  {volume} {89}},\ \bibinfo {pages} {205303} (\bibinfo {year}
  {2014})}\BibitemShut {NoStop}%
\bibitem [{\citenamefont {Wang}\ \emph {et~al.}(2018)\citenamefont {Wang},
  \citenamefont {Chernikov}, \citenamefont {Glazov}, \citenamefont {Heinz},
  \citenamefont {Marie}, \citenamefont {Amand},\ and\ \citenamefont
  {Urbaszek}}]{MX2Review}%
  \BibitemOpen
  \bibfield  {author} {\bibinfo {author} {\bibfnamefont {G.}~\bibnamefont
  {Wang}}, \bibinfo {author} {\bibfnamefont {A.}~\bibnamefont {Chernikov}},
  \bibinfo {author} {\bibfnamefont {M.~M.}\ \bibnamefont {Glazov}}, \bibinfo
  {author} {\bibfnamefont {T.~F.}\ \bibnamefont {Heinz}}, \bibinfo {author}
  {\bibfnamefont {X.}~\bibnamefont {Marie}}, \bibinfo {author} {\bibfnamefont
  {T.}~\bibnamefont {Amand}}, \ and\ \bibinfo {author} {\bibfnamefont
  {B.}~\bibnamefont {Urbaszek}},\ }\bibfield  {title} {\enquote {\bibinfo
  {title} {Colloquium: Excitons in atomically thin transition metal
  dichalcogenides},}\ }\href {\doibase 10.1103/RevModPhys.90.021001} {\bibfield
   {journal} {\bibinfo  {journal} {Rev. Mod. Phys.}\ }\textbf {\bibinfo
  {volume} {90}},\ \bibinfo {pages} {021001} (\bibinfo {year}
  {2018})}\BibitemShut {NoStop}%
\bibitem [{\citenamefont {Avdeev}\ and\ \citenamefont
  {Smirnov}(2019)}]{MX2_Avdeev}%
  \BibitemOpen
  \bibfield  {author} {\bibinfo {author} {\bibfnamefont {I.~D.}\ \bibnamefont
  {Avdeev}}\ and\ \bibinfo {author} {\bibfnamefont {D.~S.}\ \bibnamefont
  {Smirnov}},\ }\bibfield  {title} {\enquote {\bibinfo {title} {Hyperfine
  interaction in atomically thin transition metal dichalcogenides},}\ }\href
  {\doibase 10.1039/C8NA00360B} {\bibfield  {journal} {\bibinfo  {journal}
  {Nanoscale Adv.}\ }\textbf {\bibinfo {volume} {1}},\ \bibinfo {pages} {2624}
  (\bibinfo {year} {2019})}\BibitemShut {NoStop}%
\bibitem [{\citenamefont {Glazov}(2018)}]{book_Glazov}%
  \BibitemOpen
  \bibfield  {author} {\bibinfo {author} {\bibfnamefont {M.~M.}\ \bibnamefont
  {Glazov}},\ }\href@noop {} {\emph {\bibinfo {title} {Electron and Nuclear
  Spin Dynamics in Semiconductor Nanostructures}}}\ (\bibinfo  {publisher}
  {Oxford University Press, Oxford},\ \bibinfo {year} {2018})\BibitemShut
  {NoStop}%
\bibitem [{\citenamefont {Qiu}\ \emph {et~al.}(2015)\citenamefont {Qiu},
  \citenamefont {Cao},\ and\ \citenamefont {Louie}}]{PhysRevLett.115.176801}%
  \BibitemOpen
  \bibfield  {author} {\bibinfo {author} {\bibfnamefont {D.~Y.}\ \bibnamefont
  {Qiu}}, \bibinfo {author} {\bibfnamefont {T.}~\bibnamefont {Cao}}, \ and\
  \bibinfo {author} {\bibfnamefont {S.~G.}\ \bibnamefont {Louie}},\ }\bibfield
  {title} {\enquote {\bibinfo {title} {{Nonanalyticity, Valley Quantum Phases,
  and Lightlike Exciton Dispersion in Monolayer Transition Metal
  Dichalcogenides: Theory and First-Principles Calculations}},}\ }\href
  {\doibase 10.1103/PhysRevLett.115.176801} {\bibfield  {journal} {\bibinfo
  {journal} {Phys. Rev. Lett.}\ }\textbf {\bibinfo {volume} {115}},\ \bibinfo
  {pages} {176801} (\bibinfo {year} {2015})}\BibitemShut {NoStop}%
\bibitem [{\citenamefont {Peng}\ \emph {et~al.}(2019)\citenamefont {Peng},
  \citenamefont {Lo}, \citenamefont {Li}, \citenamefont {Huang}, \citenamefont
  {Chen}, \citenamefont {Lee}, \citenamefont {Yang},\ and\ \citenamefont
  {Cheng}}]{nl_exch_1}%
  \BibitemOpen
  \bibfield  {author} {\bibinfo {author} {\bibfnamefont {G.-H.}\ \bibnamefont
  {Peng}}, \bibinfo {author} {\bibfnamefont {P.-Y.}\ \bibnamefont {Lo}},
  \bibinfo {author} {\bibfnamefont {W.-H.}\ \bibnamefont {Li}}, \bibinfo
  {author} {\bibfnamefont {Y.-C.}\ \bibnamefont {Huang}}, \bibinfo {author}
  {\bibfnamefont {Y.-H.}\ \bibnamefont {Chen}}, \bibinfo {author}
  {\bibfnamefont {C.-H.}\ \bibnamefont {Lee}}, \bibinfo {author} {\bibfnamefont
  {C.-K.}\ \bibnamefont {Yang}}, \ and\ \bibinfo {author} {\bibfnamefont
  {S.-J.}\ \bibnamefont {Cheng}},\ }\bibfield  {title} {\enquote {\bibinfo
  {title} {{Distinctive Signatures of the Spin- and Momentum-Forbidden Dark
  Exciton States in the Photoluminescence of Strained WSe2 Monolayers under
  Thermalization}},}\ }\href {\doibase 10.1021/acs.nanolett.8b04786} {\bibfield
   {journal} {\bibinfo  {journal} {Nano Lett.}\ }\textbf {\bibinfo {volume}
  {19}},\ \bibinfo {pages} {2299} (\bibinfo {year} {2019})}\BibitemShut
  {NoStop}%
\bibitem [{\citenamefont {Deilmann}\ and\ \citenamefont
  {Thygesen}(2019)}]{Deilmann_2019}%
  \BibitemOpen
  \bibfield  {author} {\bibinfo {author} {\bibfnamefont {T.}~\bibnamefont
  {Deilmann}}\ and\ \bibinfo {author} {\bibfnamefont {K.~S.}\ \bibnamefont
  {Thygesen}},\ }\bibfield  {title} {\enquote {\bibinfo {title}
  {Finite-momentum exciton landscape in mono- and bilayer transition metal
  dichalcogenides},}\ }\href {\doibase 10.1088/2053-1583/ab0e1d} {\bibfield
  {journal} {\bibinfo  {journal} {2D Mater.}\ }\textbf {\bibinfo {volume}
  {6}},\ \bibinfo {pages} {035003} (\bibinfo {year} {2019})}\BibitemShut
  {NoStop}%
\bibitem [{\citenamefont {Brem}\ \emph {et~al.}(2020)\citenamefont {Brem},
  \citenamefont {Ekman}, \citenamefont {Christiansen}, \citenamefont {Katsch},
  \citenamefont {Selig}, \citenamefont {Robert}, \citenamefont {Marie},
  \citenamefont {Urbaszek}, \citenamefont {Knorr},\ and\ \citenamefont
  {Malic}}]{doi:10.1021/acs.nanolett.0c00633}%
  \BibitemOpen
  \bibfield  {author} {\bibinfo {author} {\bibfnamefont {S.}~\bibnamefont
  {Brem}}, \bibinfo {author} {\bibfnamefont {A.}~\bibnamefont {Ekman}},
  \bibinfo {author} {\bibfnamefont {D.}~\bibnamefont {Christiansen}}, \bibinfo
  {author} {\bibfnamefont {F.}~\bibnamefont {Katsch}}, \bibinfo {author}
  {\bibfnamefont {M.}~\bibnamefont {Selig}}, \bibinfo {author} {\bibfnamefont
  {C.}~\bibnamefont {Robert}}, \bibinfo {author} {\bibfnamefont
  {X.}~\bibnamefont {Marie}}, \bibinfo {author} {\bibfnamefont
  {B.}~\bibnamefont {Urbaszek}}, \bibinfo {author} {\bibfnamefont
  {A.}~\bibnamefont {Knorr}}, \ and\ \bibinfo {author} {\bibfnamefont
  {E.}~\bibnamefont {Malic}},\ }\bibfield  {title} {\enquote {\bibinfo {title}
  {{Phonon-Assisted Photoluminescence from Indirect Excitons in Monolayers of
  Transition-Metal Dichalcogenides}},}\ }\href {\doibase
  10.1021/acs.nanolett.0c00633} {\bibfield  {journal} {\bibinfo  {journal}
  {Nano Lett.}\ }\textbf {\bibinfo {volume} {20}},\ \bibinfo {pages} {2849}
  (\bibinfo {year} {2020})}\BibitemShut {NoStop}%
\bibitem [{\citenamefont {Yu}\ \emph {et~al.}(2014)\citenamefont {Yu},
  \citenamefont {Liu}, \citenamefont {Gong}, \citenamefont {Xu},\ and\
  \citenamefont {Yao}}]{Yu_exchange}%
  \BibitemOpen
  \bibfield  {author} {\bibinfo {author} {\bibfnamefont {H.}~\bibnamefont
  {Yu}}, \bibinfo {author} {\bibfnamefont {G.-B.}\ \bibnamefont {Liu}},
  \bibinfo {author} {\bibfnamefont {P.}~\bibnamefont {Gong}}, \bibinfo {author}
  {\bibfnamefont {X.}~\bibnamefont {Xu}}, \ and\ \bibinfo {author}
  {\bibfnamefont {W.}~\bibnamefont {Yao}},\ }\bibfield  {title} {\enquote
  {\bibinfo {title} {Dirac cones and Dirac saddle points of bright excitons in
  monolayer transition metal dichalcogenides},}\ }\href
  {https://doi.org/10.1038/ncomms4876} {\bibfield  {journal} {\bibinfo
  {journal} {Nat. Commun.}\ }\textbf {\bibinfo {volume} {5}},\ \bibinfo {pages}
  {3876} (\bibinfo {year} {2014})}\BibitemShut {NoStop}%
\bibitem [{\citenamefont {Wang}\ \emph {et~al.}(2015)\citenamefont {Wang},
  \citenamefont {Bouet}, \citenamefont {Glazov}, \citenamefont {Amand},
  \citenamefont {Ivchenko}, \citenamefont {Palleau}, \citenamefont {Marie},\
  and\ \citenamefont {Urbaszek}}]{Wang_2015}%
  \BibitemOpen
  \bibfield  {author} {\bibinfo {author} {\bibfnamefont {G.}~\bibnamefont
  {Wang}}, \bibinfo {author} {\bibfnamefont {L.}~\bibnamefont {Bouet}},
  \bibinfo {author} {\bibfnamefont {M.~M.}\ \bibnamefont {Glazov}}, \bibinfo
  {author} {\bibfnamefont {T.}~\bibnamefont {Amand}}, \bibinfo {author}
  {\bibfnamefont {E.~L.}\ \bibnamefont {Ivchenko}}, \bibinfo {author}
  {\bibfnamefont {E.}~\bibnamefont {Palleau}}, \bibinfo {author} {\bibfnamefont
  {X.}~\bibnamefont {Marie}}, \ and\ \bibinfo {author} {\bibfnamefont
  {B.}~\bibnamefont {Urbaszek}},\ }\bibfield  {title} {\enquote {\bibinfo
  {title} {Magneto-optics in transition metal diselenide monolayers},}\ }\href
  {\doibase 10.1088/2053-1583/2/3/034002} {\bibfield  {journal} {\bibinfo
  {journal} {2D Mater.}\ }\textbf {\bibinfo {volume} {2}},\ \bibinfo {pages}
  {034002} (\bibinfo {year} {2015})}\BibitemShut {NoStop}%
\bibitem [{\citenamefont {Durnev}\ and\ \citenamefont
  {Glazov}(2018)}]{DurnevUFN}%
  \BibitemOpen
  \bibfield  {author} {\bibinfo {author} {\bibfnamefont {M.~V.}\ \bibnamefont
  {Durnev}}\ and\ \bibinfo {author} {\bibfnamefont {M.~M.}\ \bibnamefont
  {Glazov}},\ }\bibfield  {title} {\enquote {\bibinfo {title} {Excitons and
  trions in two-dimensional semiconductors based on transition metal
  dichalcogenides},}\ }\href@noop {} {\bibfield  {journal} {\bibinfo  {journal}
  {Phys. Usp}\ }\textbf {\bibinfo {volume} {61}},\ \bibinfo {pages} {825}
  (\bibinfo {year} {2018})}\BibitemShut {NoStop}%
\bibitem [{\citenamefont {Merkulov}\ \emph {et~al.}(2002)\citenamefont
  {Merkulov}, \citenamefont {Efros},\ and\ \citenamefont {Rosen}}]{merkulov02}%
  \BibitemOpen
  \bibfield  {author} {\bibinfo {author} {\bibfnamefont {I.~A.}\ \bibnamefont
  {Merkulov}}, \bibinfo {author} {\bibfnamefont {A.~L.}\ \bibnamefont {Efros}},
  \ and\ \bibinfo {author} {\bibfnamefont {M.}~\bibnamefont {Rosen}},\
  }\bibfield  {title} {\enquote {\bibinfo {title} {Electron spin relaxation by
  nuclei in semiconductor quantum dots},}\ }\href {\doibase
  10.1103/PhysRevB.65.205309} {\bibfield  {journal} {\bibinfo  {journal} {Phys.
  Rev. B}\ }\textbf {\bibinfo {volume} {65}},\ \bibinfo {pages} {205309}
  (\bibinfo {year} {2002})}\BibitemShut {NoStop}%
\bibitem [{\citenamefont {Yu}\ \emph {et~al.}(2018)\citenamefont {Yu},
  \citenamefont {Liu},\ and\ \citenamefont {Yao}}]{Yu_2018}%
  \BibitemOpen
  \bibfield  {author} {\bibinfo {author} {\bibfnamefont {H.}~\bibnamefont
  {Yu}}, \bibinfo {author} {\bibfnamefont {G.-B.}\ \bibnamefont {Liu}}, \ and\
  \bibinfo {author} {\bibfnamefont {W.}~\bibnamefont {Yao}},\ }\bibfield
  {title} {\enquote {\bibinfo {title} {Brightened spin-triplet interlayer
  excitons and optical selection rules in van der Waals heterobilayers},}\
  }\href {\doibase 10.1088/2053-1583/aac065} {\bibfield  {journal} {\bibinfo
  {journal} {2D Mater.}\ }\textbf {\bibinfo {volume} {5}},\ \bibinfo {pages}
  {035021} (\bibinfo {year} {2018})}\BibitemShut {NoStop}%
\bibitem [{\citenamefont {Choi}\ \emph {et~al.}(2021)\citenamefont {Choi},
  \citenamefont {Florian}, \citenamefont {Steinhoff}, \citenamefont {Erben},
  \citenamefont {Tran}, \citenamefont {Kim}, \citenamefont {Sun}, \citenamefont
  {Quan}, \citenamefont {Claassen}, \citenamefont {Majumder}, \citenamefont
  {Hollingsworth}, \citenamefont {Taniguchi}, \citenamefont {Watanabe},
  \citenamefont {Ueno}, \citenamefont {Singh}, \citenamefont {Moody},
  \citenamefont {Jahnke},\ and\ \citenamefont {Li}}]{PhysRevLett.126.047401}%
  \BibitemOpen
  \bibfield  {author} {\bibinfo {author} {\bibfnamefont {J.}~\bibnamefont
  {Choi}}, \bibinfo {author} {\bibfnamefont {M.}~\bibnamefont {Florian}},
  \bibinfo {author} {\bibfnamefont {A.}~\bibnamefont {Steinhoff}}, \bibinfo
  {author} {\bibfnamefont {D.}~\bibnamefont {Erben}}, \bibinfo {author}
  {\bibfnamefont {K.}~\bibnamefont {Tran}}, \bibinfo {author} {\bibfnamefont
  {D.~S.}\ \bibnamefont {Kim}}, \bibinfo {author} {\bibfnamefont
  {L.}~\bibnamefont {Sun}}, \bibinfo {author} {\bibfnamefont {J.}~\bibnamefont
  {Quan}}, \bibinfo {author} {\bibfnamefont {R.}~\bibnamefont {Claassen}},
  \bibinfo {author} {\bibfnamefont {S.}~\bibnamefont {Majumder}}, \bibinfo
  {author} {\bibfnamefont {J.~A.}\ \bibnamefont {Hollingsworth}}, \bibinfo
  {author} {\bibfnamefont {T.}~\bibnamefont {Taniguchi}}, \bibinfo {author}
  {\bibfnamefont {K.}~\bibnamefont {Watanabe}}, \bibinfo {author}
  {\bibfnamefont {K.}~\bibnamefont {Ueno}}, \bibinfo {author} {\bibfnamefont
  {A.}~\bibnamefont {Singh}}, \bibinfo {author} {\bibfnamefont
  {G.}~\bibnamefont {Moody}}, \bibinfo {author} {\bibfnamefont
  {F.}~\bibnamefont {Jahnke}}, \ and\ \bibinfo {author} {\bibfnamefont
  {X.}~\bibnamefont {Li}},\ }\bibfield  {title} {\enquote {\bibinfo {title}
  {Twist Angle-Dependent Interlayer Exciton Lifetimes in van der Waals
  Heterostructures},}\ }\href {\doibase 10.1103/PhysRevLett.126.047401}
  {\bibfield  {journal} {\bibinfo  {journal} {Phys. Rev. Lett.}\ }\textbf
  {\bibinfo {volume} {126}},\ \bibinfo {pages} {047401} (\bibinfo {year}
  {2021})}\BibitemShut {NoStop}%
\bibitem [{\citenamefont {Jiang}\ \emph {et~al.}(2018)\citenamefont {Jiang},
  \citenamefont {Xu}, \citenamefont {Rasmita}, \citenamefont {Huang},
  \citenamefont {Li}, \citenamefont {Xiong},\ and\ \citenamefont
  {Gao}}]{Jiang2018}%
  \BibitemOpen
  \bibfield  {author} {\bibinfo {author} {\bibfnamefont {C.}~\bibnamefont
  {Jiang}}, \bibinfo {author} {\bibfnamefont {W.}~\bibnamefont {Xu}}, \bibinfo
  {author} {\bibfnamefont {A.}~\bibnamefont {Rasmita}}, \bibinfo {author}
  {\bibfnamefont {Z.}~\bibnamefont {Huang}}, \bibinfo {author} {\bibfnamefont
  {K.}~\bibnamefont {Li}}, \bibinfo {author} {\bibfnamefont {Q.}~\bibnamefont
  {Xiong}}, \ and\ \bibinfo {author} {\bibfnamefont {W.-b.}\ \bibnamefont
  {Gao}},\ }\bibfield  {title} {\enquote {\bibinfo {title} {Microsecond
  dark-exciton valley polarization memory in two-dimensional
  heterostructures},}\ }\href {https://doi.org/10.1038/s41467-018-03174-3}
  {\bibfield  {journal} {\bibinfo  {journal} {Nat. Commun.}\ }\textbf {\bibinfo
  {volume} {9}},\ \bibinfo {pages} {753} (\bibinfo {year} {2018})}\BibitemShut
  {NoStop}%
\bibitem [{\citenamefont {{Wo\ifmmode \acute{z}\else {\'z}\fi{}niak}}\ \emph
  {et~al.}(2020)\citenamefont {{Wo\ifmmode \acute{z}\else {\'z}\fi{}niak}},
  \citenamefont {{Faria Junior}}, \citenamefont {Seifert}, \citenamefont
  {Chaves},\ and\ \citenamefont {Kunstmann}}]{PhysRevB.101.235408}%
  \BibitemOpen
  \bibfield  {author} {\bibinfo {author} {\bibfnamefont {T.}~\bibnamefont
  {{Wo\ifmmode \acute{z}\else {\'z}\fi{}niak}}}, \bibinfo {author}
  {\bibfnamefont {P.~E.}\ \bibnamefont {{Faria Junior}}}, \bibinfo {author}
  {\bibfnamefont {G.}~\bibnamefont {Seifert}}, \bibinfo {author} {\bibfnamefont
  {A.}~\bibnamefont {Chaves}}, \ and\ \bibinfo {author} {\bibfnamefont
  {J.}~\bibnamefont {Kunstmann}},\ }\bibfield  {title} {\enquote {\bibinfo
  {title} {Exciton $g$ factors of van der Waals heterostructures from
  first-principles calculations},}\ }\href {\doibase
  10.1103/PhysRevB.101.235408} {\bibfield  {journal} {\bibinfo  {journal}
  {Phys. Rev. B}\ }\textbf {\bibinfo {volume} {101}},\ \bibinfo {pages}
  {235408} (\bibinfo {year} {2020})}\BibitemShut {NoStop}%
\bibitem [{\citenamefont {Xuan}\ and\ \citenamefont
  {Quek}(2020)}]{PhysRevResearch.2.033256}%
  \BibitemOpen
  \bibfield  {author} {\bibinfo {author} {\bibfnamefont {F.}~\bibnamefont
  {Xuan}}\ and\ \bibinfo {author} {\bibfnamefont {S.~Y.}\ \bibnamefont
  {Quek}},\ }\bibfield  {title} {\enquote {\bibinfo {title} {Valley Zeeman
  effect and Landau levels in two-dimensional transition metal
  dichalcogenides},}\ }\href {\doibase 10.1103/PhysRevResearch.2.033256}
  {\bibfield  {journal} {\bibinfo  {journal} {Phys. Rev. Research}\ }\textbf
  {\bibinfo {volume} {2}},\ \bibinfo {pages} {033256} (\bibinfo {year}
  {2020})}\BibitemShut {NoStop}%
\bibitem [{\citenamefont {Deilmann}\ \emph {et~al.}(2020)\citenamefont
  {Deilmann}, \citenamefont {Kr{\"u}ger},\ and\ \citenamefont
  {Rohlfing}}]{PhysRevLett.124.226402}%
  \BibitemOpen
  \bibfield  {author} {\bibinfo {author} {\bibfnamefont {T.}~\bibnamefont
  {Deilmann}}, \bibinfo {author} {\bibfnamefont {P.}~\bibnamefont
  {Kr{\"u}ger}}, \ and\ \bibinfo {author} {\bibfnamefont {M.}~\bibnamefont
  {Rohlfing}},\ }\bibfield  {title} {\enquote {\bibinfo {title} {Ab Initio
  Studies of Exciton $g$ Factors: Monolayer Transition Metal Dichalcogenides in
  Magnetic Fields},}\ }\href {\doibase 10.1103/PhysRevLett.124.226402}
  {\bibfield  {journal} {\bibinfo  {journal} {Phys. Rev. Lett.}\ }\textbf
  {\bibinfo {volume} {124}},\ \bibinfo {pages} {226402} (\bibinfo {year}
  {2020})}\BibitemShut {NoStop}%
\bibitem [{Note1()}]{Note1}%
  \BibitemOpen
  \bibinfo {note} {This assumption is valid for $\gamma _e\gg G\gg (|g_e|\mu
  _B\Delta _B)/\hbar $.}\BibitemShut {Stop}%
\bibitem [{\citenamefont {Hsu}\ \emph {et~al.}(2015)\citenamefont {Hsu},
  \citenamefont {Chen}, \citenamefont {Chen}, \citenamefont {Liu},
  \citenamefont {Hou}, \citenamefont {Li},\ and\ \citenamefont
  {Chang}}]{Hsu2015}%
  \BibitemOpen
  \bibfield  {author} {\bibinfo {author} {\bibfnamefont {W.-T.}\ \bibnamefont
  {Hsu}}, \bibinfo {author} {\bibfnamefont {Y.-L.}\ \bibnamefont {Chen}},
  \bibinfo {author} {\bibfnamefont {C.-H.}\ \bibnamefont {Chen}}, \bibinfo
  {author} {\bibfnamefont {P.-S.}\ \bibnamefont {Liu}}, \bibinfo {author}
  {\bibfnamefont {T.-H.}\ \bibnamefont {Hou}}, \bibinfo {author} {\bibfnamefont
  {L.-J.}\ \bibnamefont {Li}}, \ and\ \bibinfo {author} {\bibfnamefont {W.-H.}\
  \bibnamefont {Chang}},\ }\bibfield  {title} {\enquote {\bibinfo {title}
  {Optically initialized robust valley-polarized holes in monolayer WSe$_2$},}\
  }\href@noop {} {\bibfield  {journal} {\bibinfo  {journal} {Nat. commun.}\
  }\textbf {\bibinfo {volume} {6}},\ \bibinfo {pages} {8963} (\bibinfo {year}
  {2015})}\BibitemShut {NoStop}%
\bibitem [{\citenamefont {Yang}\ \emph {et~al.}(2015)\citenamefont {Yang},
  \citenamefont {Sinitsyn}, \citenamefont {Chen}, \citenamefont {Yuan},
  \citenamefont {Zhang}, \citenamefont {Lou},\ and\ \citenamefont
  {Crooker}}]{Yang2015nat}%
  \BibitemOpen
  \bibfield  {author} {\bibinfo {author} {\bibfnamefont {L.}~\bibnamefont
  {Yang}}, \bibinfo {author} {\bibfnamefont {N.~A.}\ \bibnamefont {Sinitsyn}},
  \bibinfo {author} {\bibfnamefont {W.}~\bibnamefont {Chen}}, \bibinfo {author}
  {\bibfnamefont {J.}~\bibnamefont {Yuan}}, \bibinfo {author} {\bibfnamefont
  {J.}~\bibnamefont {Zhang}}, \bibinfo {author} {\bibfnamefont
  {J.}~\bibnamefont {Lou}}, \ and\ \bibinfo {author} {\bibfnamefont {S.~A.}\
  \bibnamefont {Crooker}},\ }\bibfield  {title} {\enquote {\bibinfo {title}
  {Long-lived nanosecond spin relaxation and spin coherence of electrons in
  monolayer MoS$_2$ and WS$_2$},}\ }\href@noop {} {\bibfield  {journal}
  {\bibinfo  {journal} {Nat. Phys.}\ }\textbf {\bibinfo {volume} {11}},\
  \bibinfo {pages} {830} (\bibinfo {year} {2015})}\BibitemShut {NoStop}%
\bibitem [{\citenamefont {Dey}\ \emph {et~al.}(2017)\citenamefont {Dey},
  \citenamefont {Yang}, \citenamefont {Robert}, \citenamefont {Wang},
  \citenamefont {Urbaszek}, \citenamefont {Marie},\ and\ \citenamefont
  {Crooker}}]{Dey2017}%
  \BibitemOpen
  \bibfield  {author} {\bibinfo {author} {\bibfnamefont {P.}~\bibnamefont
  {Dey}}, \bibinfo {author} {\bibfnamefont {L.}~\bibnamefont {Yang}}, \bibinfo
  {author} {\bibfnamefont {C.}~\bibnamefont {Robert}}, \bibinfo {author}
  {\bibfnamefont {G.}~\bibnamefont {Wang}}, \bibinfo {author} {\bibfnamefont
  {B.}~\bibnamefont {Urbaszek}}, \bibinfo {author} {\bibfnamefont
  {X.}~\bibnamefont {Marie}}, \ and\ \bibinfo {author} {\bibfnamefont {S.~A.}\
  \bibnamefont {Crooker}},\ }\bibfield  {title} {\enquote {\bibinfo {title}
  {Gate-Controlled Spin-Valley Locking of Resident Carriers in
  ${\mathrm{WSe}}_{2}$ Monolayers},}\ }\href {\doibase
  10.1103/PhysRevLett.119.137401} {\bibfield  {journal} {\bibinfo  {journal}
  {Phys. Rev. Lett.}\ }\textbf {\bibinfo {volume} {119}},\ \bibinfo {pages}
  {137401} (\bibinfo {year} {2017})}\BibitemShut {NoStop}%
\bibitem [{\citenamefont {Jin}\ \emph {et~al.}(2018)\citenamefont {Jin},
  \citenamefont {Kim}, \citenamefont {Utama}, \citenamefont {Regan},
  \citenamefont {Kleemann}, \citenamefont {Cai}, \citenamefont {Shen},
  \citenamefont {Shinner}, \citenamefont {Sengupta}, \citenamefont {Watanabe},
  \citenamefont {Taniguchi}, \citenamefont {Tongay}, \citenamefont {Zettl},\
  and\ \citenamefont {Wang}}]{Jin893}%
  \BibitemOpen
  \bibfield  {author} {\bibinfo {author} {\bibfnamefont {C.}~\bibnamefont
  {Jin}}, \bibinfo {author} {\bibfnamefont {J.}~\bibnamefont {Kim}}, \bibinfo
  {author} {\bibfnamefont {M.~I.~B.}\ \bibnamefont {Utama}}, \bibinfo {author}
  {\bibfnamefont {E.~C.}\ \bibnamefont {Regan}}, \bibinfo {author}
  {\bibfnamefont {H.}~\bibnamefont {Kleemann}}, \bibinfo {author}
  {\bibfnamefont {H.}~\bibnamefont {Cai}}, \bibinfo {author} {\bibfnamefont
  {Y.}~\bibnamefont {Shen}}, \bibinfo {author} {\bibfnamefont {M.~J.}\
  \bibnamefont {Shinner}}, \bibinfo {author} {\bibfnamefont {A.}~\bibnamefont
  {Sengupta}}, \bibinfo {author} {\bibfnamefont {K.}~\bibnamefont {Watanabe}},
  \bibinfo {author} {\bibfnamefont {T.}~\bibnamefont {Taniguchi}}, \bibinfo
  {author} {\bibfnamefont {S.}~\bibnamefont {Tongay}}, \bibinfo {author}
  {\bibfnamefont {A.}~\bibnamefont {Zettl}}, \ and\ \bibinfo {author}
  {\bibfnamefont {F.}~\bibnamefont {Wang}},\ }\bibfield  {title} {\enquote
  {\bibinfo {title} {Imaging of pure spin-valley diffusion current in
  WS$_2$-WSe$_2$ heterostructures},}\ }\href {\doibase 10.1126/science.aao3503}
  {\bibfield  {journal} {\bibinfo  {journal} {Science}\ }\textbf {\bibinfo
  {volume} {360}},\ \bibinfo {pages} {893} (\bibinfo {year}
  {2018})}\BibitemShut {NoStop}%
\bibitem [{\citenamefont {Ivchenko}(2018)}]{Ivchenko2018}%
  \BibitemOpen
  \bibfield  {author} {\bibinfo {author} {\bibfnamefont {E.~L.}\ \bibnamefont
  {Ivchenko}},\ }\bibfield  {title} {\enquote {\bibinfo {title} {Magnetic
  Circular Polarization of Exciton Photoluminescence},}\ }\href {\doibase
  10.1134/S1063783418080127} {\bibfield  {journal} {\bibinfo  {journal} {Phys.
  Solid State}\ }\textbf {\bibinfo {volume} {60}},\ \bibinfo {pages} {1514}
  (\bibinfo {year} {2018})}\BibitemShut {NoStop}%
\bibitem [{\citenamefont {Smirnov}\ \emph
  {et~al.}(2020{\natexlab{a}})\citenamefont {Smirnov}, \citenamefont
  {Shamirzaev}, \citenamefont {Yakovlev},\ and\ \citenamefont
  {Bayer}}]{PhysRevLett.125.156801}%
  \BibitemOpen
  \bibfield  {author} {\bibinfo {author} {\bibfnamefont {D.~S.}\ \bibnamefont
  {Smirnov}}, \bibinfo {author} {\bibfnamefont {T.~S.}\ \bibnamefont
  {Shamirzaev}}, \bibinfo {author} {\bibfnamefont {D.~R.}\ \bibnamefont
  {Yakovlev}}, \ and\ \bibinfo {author} {\bibfnamefont {M.}~\bibnamefont
  {Bayer}},\ }\bibfield  {title} {\enquote {\bibinfo {title} {Dynamic
  Polarization of Electron Spins Interacting with Nuclei in Semiconductor
  Nanostructures},}\ }\href {\doibase 10.1103/PhysRevLett.125.156801}
  {\bibfield  {journal} {\bibinfo  {journal} {Phys. Rev. Lett.}\ }\textbf
  {\bibinfo {volume} {125}},\ \bibinfo {pages} {156801} (\bibinfo {year}
  {2020}{\natexlab{a}})}\BibitemShut {NoStop}%
\bibitem [{\citenamefont {Shamirzaev}\ \emph {et~al.}(2021)\citenamefont
  {Shamirzaev}, \citenamefont {Shumilin}, \citenamefont {Smirnov},
  \citenamefont {Rautert}, \citenamefont {Yakovlev},\ and\ \citenamefont
  {Bayer}}]{shamirzaev2021dynamic}%
  \BibitemOpen
  \bibfield  {author} {\bibinfo {author} {\bibfnamefont {T.~S.}\ \bibnamefont
  {Shamirzaev}}, \bibinfo {author} {\bibfnamefont {A.~V.}\ \bibnamefont
  {Shumilin}}, \bibinfo {author} {\bibfnamefont {D.~S.}\ \bibnamefont
  {Smirnov}}, \bibinfo {author} {\bibfnamefont {J.}~\bibnamefont {Rautert}},
  \bibinfo {author} {\bibfnamefont {D.~R.}\ \bibnamefont {Yakovlev}}, \ and\
  \bibinfo {author} {\bibfnamefont {M.}~\bibnamefont {Bayer}},\ }\href@noop {}
  {\enquote {\bibinfo {title} {Dynamic polarization of electron spins in
  indirect band gap (In,Al)As/AlAs quantum dots in weak magnetic field:
  experiment and theory},}\ } (\bibinfo {year} {2021}),\ \Eprint
  {http://arxiv.org/abs/2106.00044} {arXiv:2106.00044} \BibitemShut {NoStop}%
\bibitem [{\citenamefont {Smirnov}\ \emph
  {et~al.}(2020{\natexlab{b}})\citenamefont {Smirnov}, \citenamefont {Zhukov},
  \citenamefont {Yakovlev}, \citenamefont {Kirstein}, \citenamefont {Bayer},\
  and\ \citenamefont {Greilich}}]{PRC}%
  \BibitemOpen
  \bibfield  {author} {\bibinfo {author} {\bibfnamefont {D.~S.}\ \bibnamefont
  {Smirnov}}, \bibinfo {author} {\bibfnamefont {E.~A.}\ \bibnamefont {Zhukov}},
  \bibinfo {author} {\bibfnamefont {D.~R.}\ \bibnamefont {Yakovlev}}, \bibinfo
  {author} {\bibfnamefont {E.}~\bibnamefont {Kirstein}}, \bibinfo {author}
  {\bibfnamefont {M.}~\bibnamefont {Bayer}}, \ and\ \bibinfo {author}
  {\bibfnamefont {A.}~\bibnamefont {Greilich}},\ }\bibfield  {title} {\enquote
  {\bibinfo {title} {Spin polarization recovery and Hanle effect for charge
  carriers interacting with nuclear spins in semiconductors},}\ }\href
  {\doibase 10.1103/PhysRevB.102.235413} {\bibfield  {journal} {\bibinfo
  {journal} {Phys. Rev. B}\ }\textbf {\bibinfo {volume} {102}},\ \bibinfo
  {pages} {235413} (\bibinfo {year} {2020}{\natexlab{b}})}\BibitemShut
  {NoStop}%
\bibitem [{\citenamefont {{Smole\ifmmode \acute{n}\else {\'n}\fi{}ski}}\ \emph
  {et~al.}(2016)\citenamefont {{Smole\ifmmode \acute{n}\else {\'n}\fi{}ski}},
  \citenamefont {Goryca}, \citenamefont {Koperski}, \citenamefont {Faugeras},
  \citenamefont {Kazimierczuk}, \citenamefont {Bogucki}, \citenamefont
  {Nogajewski}, \citenamefont {Kossacki},\ and\ \citenamefont
  {Potemski}}]{PhysRevX.6.021024}%
  \BibitemOpen
  \bibfield  {author} {\bibinfo {author} {\bibfnamefont {T.}~\bibnamefont
  {{Smole\ifmmode \acute{n}\else {\'n}\fi{}ski}}}, \bibinfo {author}
  {\bibfnamefont {M.}~\bibnamefont {Goryca}}, \bibinfo {author} {\bibfnamefont
  {M.}~\bibnamefont {Koperski}}, \bibinfo {author} {\bibfnamefont
  {C.}~\bibnamefont {Faugeras}}, \bibinfo {author} {\bibfnamefont
  {T.}~\bibnamefont {Kazimierczuk}}, \bibinfo {author} {\bibfnamefont
  {A.}~\bibnamefont {Bogucki}}, \bibinfo {author} {\bibfnamefont
  {K.}~\bibnamefont {Nogajewski}}, \bibinfo {author} {\bibfnamefont
  {P.}~\bibnamefont {Kossacki}}, \ and\ \bibinfo {author} {\bibfnamefont
  {M.}~\bibnamefont {Potemski}},\ }\bibfield  {title} {\enquote {\bibinfo
  {title} {Tuning Valley Polarization in a ${\mathrm{WSe}}_{2}$ Monolayer with
  a Tiny Magnetic Field},}\ }\href {\doibase 10.1103/PhysRevX.6.021024}
  {\bibfield  {journal} {\bibinfo  {journal} {Phys. Rev. X}\ }\textbf {\bibinfo
  {volume} {6}},\ \bibinfo {pages} {021024} (\bibinfo {year}
  {2016})}\BibitemShut {NoStop}%
\bibitem [{\citenamefont {Li}\ \emph {et~al.}(2021)\citenamefont {Li},
  \citenamefont {Goryca}, \citenamefont {Yumigeta}, \citenamefont {Li},
  \citenamefont {Tongay},\ and\ \citenamefont
  {Crooker}}]{PhysRevMaterials.5.044001}%
  \BibitemOpen
  \bibfield  {author} {\bibinfo {author} {\bibfnamefont {J.}~\bibnamefont
  {Li}}, \bibinfo {author} {\bibfnamefont {M.}~\bibnamefont {Goryca}}, \bibinfo
  {author} {\bibfnamefont {K.}~\bibnamefont {Yumigeta}}, \bibinfo {author}
  {\bibfnamefont {H.}~\bibnamefont {Li}}, \bibinfo {author} {\bibfnamefont
  {S.}~\bibnamefont {Tongay}}, \ and\ \bibinfo {author} {\bibfnamefont {S.~A.}\
  \bibnamefont {Crooker}},\ }\bibfield  {title} {\enquote {\bibinfo {title}
  {Valley relaxation of resident electrons and holes in a monolayer
  semiconductor: Dependence on carrier density and the role of
  substrate-induced disorder},}\ }\href {\doibase
  10.1103/PhysRevMaterials.5.044001} {\bibfield  {journal} {\bibinfo  {journal}
  {Phys. Rev. Materials}\ }\textbf {\bibinfo {volume} {5}},\ \bibinfo {pages}
  {044001} (\bibinfo {year} {2021})}\BibitemShut {NoStop}%
\end{thebibliography}
\end{document}